\documentclass[preprint]{emulateapj}

\usepackage{apjfonts}
\usepackage{epsfig}
\usepackage{natbib}
\usepackage{float}
\usepackage{amsmath}
\usepackage{rotating}

\def\Msun{$M_\odot$}
\def\ser{S{\'e}rsic}
\def\mus{{{ $\mu m $}}}

\def \apjs {ApJS}
\def \apjl {ApJL}
\def \aap {A\&A}

\def \mus{{$\mu {\rm m}$ }}
\def \micron{{$\mu {\rm m}$}}
\def \s4g{{S$^{4}$G}}

\def \sig{{${\sigma}$}}

\shorttitle{Disk break of barred galaxies in {\s4g}}
\shortauthors{Kim et al.}

\begin{document}
\title{Unveiling the Structure of Barred Galaxies at 3.6 {\micron} with the Spitzer Survey of Stellar Structure in Galaxies (S$^4$G): I. Disk Breaks}

\author{
Taehyun Kim\altaffilmark{1,2,3,4},
Dimitri A. Gadotti\altaffilmark{2},
Kartik Sheth\altaffilmark{3},
E. Athanassoula\altaffilmark{5},
Albert Bosma\altaffilmark{5},
Myung~Gyoon Lee\altaffilmark{1},
Barry~F. Madore\altaffilmark{4},
Bruce Elmegreen\altaffilmark{6},
Johan H. Knapen\altaffilmark{7,8},
Dennis Zaritsky\altaffilmark{9},
Luis C. Ho\altaffilmark{4},
S\'ebastien Comer\'on\altaffilmark{10,11},
Benne Holwerda\altaffilmark{12},
Joannah L. Hinz\altaffilmark{13},
Juan-Carlos Mu\~noz-Mateos\altaffilmark{2,3},
Mauricio~Cisternas\altaffilmark{7,8},
Santiago Erroz-Ferrer\altaffilmark{7,8},
Ron~Buta\altaffilmark{14},
Eija Laurikainen\altaffilmark{10,11},
Heikki Salo\altaffilmark{10},
Jarkko~Laine\altaffilmark{10},
Kar\'in~Men\'endez-Delmestre\altaffilmark{15},
Michael W. Regan\altaffilmark{16},
Bonita de Swardt\altaffilmark{17},
Armando Gil de Paz\altaffilmark{18},
Mark Seibert\altaffilmark{4},
Trisha Mizusawa\altaffilmark{3,19}
}
\altaffiltext{1}{Astronomy Program, Department of Physics and Astronomy, Seoul National University, Seoul 151-742, Korea; thkim@astro.snu.ac.kr}
\altaffiltext{2}{European Southern Observatory, Casilla 19001, Santiago 19, Chile}
\altaffiltext{3}{National Radio Astronomy Observatory/NAASC, 520 Edgemont Road, Charlottesville, VA 22903, USA}
\altaffiltext{4}{The Observatories of the Carnegie Institution of Washington, 813 Santa Barbara Street, Pasadena, CA 91101, USA}
\altaffiltext{5}{Aix Marseille Universit{\'e}, CNRS, LAM (Laboratoire d'Astrophysique de Marseille) UMR 7326, 13388 Marseille, France}
\altaffiltext{6}{IBM Research Division, T.J. Watson Research Center, Yorktown Hts., NY 10598, USA}
\altaffiltext{7}{Instituto de Astrof\'\i sica de Canarias, E-38200 La Laguna, Tenerife, Spain}
\altaffiltext{8}{Departamento de Astrof\'\i sica, Universidad de La Laguna, E-38205 La Laguna, Tenerife, Spain}
\altaffiltext{9}{University of Arizona, 933 N. Cherry Ave, Tucson, AZ 85721, USA}
\altaffiltext{10}{Division of Astronomy, Department of Physical Sciences, University of Oulu, Oulu, FIN-90014, Finland}
\altaffiltext{11}{Finnish Centre of Astronomy with ESO (FINCA), University of Turku, V{\"a}is{\"a}l{\"a}ntie 20, FI-21500, Piikki{\"o}, Finland}
\altaffiltext{12}{European Space Agency, ESTEC, Keplerlaan 1, 2200-AG, Noordwijk, The Netherlands}
\altaffiltext{13}{MMTO, University of Arizona, 933 North Cherry Avenue, Tucson, AZ 85721, USA}
\altaffiltext{14}{Department of Physics and Astronomy, University of Alabama, Box 870324, Tuscaloosa, AL 35487, USA}
\altaffiltext{15}{Universidade Federal do Rio de Janeiro, Observat{\'o}rio do Valongo, Ladeira Pedro Ant{\^{o}}nio, 43, CEP 20080-090, Rio de Janeiro, Brazil}
\altaffiltext{16}{Space Telescope Science Institute, 3700 San Martin Drive, Baltimore, MD 21218, USA}
\altaffiltext{17}{South African Astronomical Observatory, Observatory, 7935 Cape Town, South Africa}
\altaffiltext{18}{Departamento de Astrof\'\i sica, Universidad Complutense de Madrid, Madrid 28040, Spain}
\altaffiltext{19}{Florida Institute of Technology, Melbourne, FL 32901, USA}

\begin{abstract}

We have performed two-dimensional multicomponent decomposition of 144 local barred spiral galaxies using 3.6 \mus images from the {\it Spitzer} Survey of Stellar Structure in Galaxies. Our model fit includes up to four components (bulge, disk, bar, and a point source) and, most importantly, takes into account disk breaks. We find that ignoring the disk break and using a single disk scale length in the model fit for Type II (down-bending) disk galaxies can lead to differences of 40\% in the disk scale length, 10\% in bulge-to-total luminosity ratio (B/T), and 25\% in bar-to-total luminosity ratios. We find that for galaxies with B/T $\geq$ 0.1, the break radius to bar radius, $r_{\rm br}/R_{\rm bar}$, varies between 1 and 3, but as a function of B/T the ratio remains roughly constant. This suggests that in bulge-dominated galaxies the disk break is likely related to the outer Lindblad Resonance (OLR) of the bar, and thus moves outwards as the bar grows. For galaxies with small bulges, B/T $<$ 0.1, $r_{\rm br}/R_{\rm bar}$ spans a wide range from 1 to 6. This suggests that the mechanism that produces the break in these galaxies may be different from that in galaxies with more massive bulges. Consistent with previous studies, we conclude that disk breaks in galaxies with small bulges may originate from bar resonances that may be also coupled with the spiral arms, or be related to star formation thresholds.

\end{abstract}

\keywords{galaxies: evolution -- galaxies: formation -- galaxies: fundamental parameters -- galaxies: photometry -- galaxies: spiral --  galaxies: structure}

\section{INTRODUCTION}

The structural components of a galaxy evolve over cosmic time. Their present day properties provide important clues to their formation and evolutionary history and provide strong constraints for cosmological simulations seeking to reproduce realistic galaxy disks.  
One key tool in quantifying the structure of disks is its radial surface brightness profile. 
Typically galaxy disks have been modeled with a single exponential function (Type I, \citealt{freeman_70}).
However, \citet{vanderkruit_79} found that some edge-on galaxies showed sharp edges in their radial surface brightness profile with a truncation radius of a few disk scale lengths (\citealt{vanderkruit_81a}).
A number of recent studies of face-on galaxies \citep[]{pohlen_02, erwin_05_anti, pohlen_06, erwin_08, gutierrez_11, maltby_12a, munoz_mateos_13} and edge-on galaxies \citep[]{comeron_12, martin_navarro_12} have found that instead of a sharp truncation there is a change in the slope of the radial surface brightness profile; the light profile is better modeled with two components - an inner and outer disk with different scale lengths. 
The transition between the two profiles is often referred to as the ``break'' in the profile.  
Galaxies with a down-bending light profile with a shallower inner disk and a steeper outer disk are referred to as Type II and those with an up-bending profile with a steeper inner disk and a shallower outer disk are referred to as Type III (see, e.g., \citealt{pohlen_02,erwin_05_anti,hunter_06,pohlen_06}).  

For NGC 4244 and NGC 7793, breaks are found in all stellar populations at the same position (\citealt{dejong_07, radburn_smith_12}), though changes in the slope at the break are different among stellar populations, with a milder transition in older stellar populations (\citealt{radburn_smith_12}). Indeed, numerical simulations (\citealt{roskar_08a, sanchez_blazquez_09}) expect that through radial stellar migrations older stars will show shallower radial profiles, because older populations are subject to scattering for longer. This explains the U-shaped color profile, with a minimum at the break radius, found in face-on Type II galaxies (\citealt{bakos_08}). However, \citet{martin_navarro_12} could not confirm this kind of color profiles in edge-on galaxies, presumably because of the increased role of dust extinction in edge-on galaxies
    
A significant fraction of disks have double exponential profiles. \citet{gutierrez_11} gathered data from two studies (\citealt{pohlen_06}, \citealt{erwin_08}) and reported that 21, 50, and 38\% of disks (Hubble types from S0 to Sm) have a Type I, II, and III profiles respectively.
Furthermore, they found that 8\% of galaxies show two breaks with composite profile, Type II$+$III (the total percentage is higher than 100\% because they count composite Type II$+$III profiles twice in their Type II and Type III fractions). They also found a trend that Type II disks increase from 25\% in S0 galaxies to 80\% in Sd and Sm galaxies, while Type I disks decrease from 30\% in early type spirals to 10\% in late type spirals.  

Given that such a high fraction of disks has a break in their radial profile, it is critical to account for the disk break when decomposing and modeling galaxies.  We note that in the literature the terms ``break'' and ``truncation'' are  often used interchangeably to refer the location at which the radial light profile changes its slope. However, we will use the term ``break'' to refer the feature that occurs well inside the disk that we focus on for this study and use the term ``truncation'' to refer the characteristic feature in edge-on disks that occurs farther out (see \citealt{martin_navarro_12} for details on breaks and truncations). 
This approach of considering breaks and truncations as fundamentally different entities is supported by recent work by S. P. C. Peters et al. (2013, in prep), who study deep imaging from the Stripe82 region of SDSS for 22 nearby, face-on spiral galaxies. They find that breaks occur in almost all galaxies at $\mu_{\rm r} \sim$ 23 mag arcsec$^{-2}$, but truncations occur in only 23\% (5/22) of their galaxies at $\mu_{\rm r} \sim $27 mag arcsec$^{-2}$ and are only visible in galaxies where a stellar halo is not detected. 
Due to projection effects, the truncation occurs at much higher surface brightness levels in edge-on galaxies, which explains why they were seen in such galaxies already decades ago (e.g., \citealt{vanderkruit_79}).

\subsection{The Origin of Breaks}
Several theories have been proposed to explain the physical origin of disk breaks. 
Van der Kruit (\citeyear{vanderkruit_87}) suggested that the break occurs at the radius of the maximum angular momentum of the protogalactic cloud in the course of galaxy formation. \citet{kennicutt_89} proposed that breaks can arise where the density of gas in the disk falls below a critical threshold beyond which stars cannot form.  
\citet{elmegreen_94} and \citet{schaye_04} suggested that the break happens where there is a phase transition of the interstellar medium between cool and warm phases. \citet{elmegreen_06} suggested that in outer disks, where the average gas column density is below the critical threshold, stars may still form from turbulent compression and other dynamical processes. Thus even in a galaxy with a single exponential gas distribution the different star formation modes can lead to a double exponential with a shallower inner disk and a steeper outer disk. 
Using N-body simulations, \citet{debattista_06} proposed that breaks can occur as a result of angular momentum redistribution induced by non-axisymmetric structures such as bars. 
\citet{foyle_08} built galaxy models using a N-body/smoothed particle hydrodynamics code and let them evolve without any interaction or gas accretion. They showed that, as a result of angular momentum redistribution, purely exponential disks evolved to develop a break. In their simulations the density profiles of inner disks evolved remarkably, while the density profile of outer disks remained relatively stable over time.

For the up-bending (Type III) disk profiles \citet{laurikainen_01} demonstrated that galaxies encountering a less massive companion (e.g., M51-like systems) developed an up-bending profile due to the stripping of stars and gas from the larger, more massive disk during the interaction. In their N-body simulations they found that these up-bending profiles remain visible for several Gyr after the passage. \citet{younger_07} showed that minor mergers could produce an up-bending profile -- gas inflows driven by mergers accumulate mass in the central regions while the outer disk expands as angular momentum is transferred outwards. \citet{kazantzidis_09} suggested that up-bending profiles could be induced by the dynamical response of thin galactic disks to the accretion of cold dark matter substructures/sub-halos.  Genuine up-bending may not be common and many may be artifacts from a superposition of a thin and a thick disk with different scale lengths (\citealt{comeron_12}). \citet{maltby_12b} concluded that 15\% of up-bending profiles are due to an extended spheroidal component.  

Studies have shown that once disk galaxies are massive enough and rotationally supported, the bar instability develops relatively fast within a few hundred Myr (e.g., \citealt{hohl_71,kalnajs_72, ostriker_73, sheth_12}). Two thirds of local disk galaxies are barred (e.g., \citealt{eskridge_00,menendez_delmestre_07, sheth_08}) and bars fractions are estimated up to z$\sim$0.8 finding that bars in more massive, early type, red and bulge dominated systems formed earlier than those in less massive, late type and blue systems (\citealt{sheth_08, cameron_10}).
Although not all barred galaxies exhibit a break, the occurrence of a break is closely linked to bars (\citealt{debattista_06,foyle_08}) as bars drive angular momentum redistribution. Therefore, we focus on barred galaxies in this study. Detailed study of unbarred galaxies will be the subject of a future study.  

\subsection{Decomposing Galaxies}
Decomposing galaxy images into their different structural components has become a major tool for studying the formation and evolution of galaxies recently. 
From the pioneering works of, e.g., \citet{dejong_95_t}, the two-dimensional (2D) decomposition technique has evolved considerably with studies by many authors \citep[including][]{marleau_98,khoroshahi_00,mollenhoff_01,donofrio_01b,peng_02, peng_10, desouza_04, laurikainen_04_bar, laurikainen_05,laurikainen_10, pignatelli_06, allen_06, haussler_07, huertas_company_07, gadotti_08, gadotti_09, durbala_08}. 
The main products from these decomposition algorithms are physical parameters for each structural component such as the effective radius of the bulge, the scale length of the disk, the ellipticity of the bar, the bulge-to-total ratio etc.
These studies have revealed significant complexity in structures and have lead to refinements in the fitting techniques as well as our understanding of galaxy evolution. Examples of such changes are the fitting of both light profiles and 2D shapes of bulges and elliptical galaxies. Light profiles are fitted with S\'ersic functions with a free S\'ersic index $n$ rather than a (n=4) de Vaucouleurs function \citep[see, e.g.,][]{caon_93,andredakis_95,laurikainen_07,gadotti_09}, or, 
multicomponent fitting of S0 galaxies with lenses (\citealt{laurikainen_05, laurikainen_07, laurikainen_09}), or, more recently, the suggestion that a majority of nearby elliptical galaxies contain not one but as many as three sub-components (e.g., \citealt{huang_13}).  
 
\citet{graham_01} demonstrated that fitting bulges with the assumption that $n=4$ yielded higher luminosities and larger sizes than if $n$ is actually smaller than 4, and lower luminosities and smaller sizes than if $n$ is greater than 4 (also see \citealt{kim_12}). A similar development is noted in the modeling of barred galaxies -- excluding a bar from the fit leads to very poor estimates for the bulge and disk parameters \citep[see, e.g.,][]{laurikainen_05,gadotti_08}.

To our knowledge, accounting for disk breaks has not yet been implemented in 2D decompositions of any large samples of galaxies. The effects of fitting single exponentials, when a double exponential is a better and truer representation of the light profile, can in particular cases be small \citep[see][]{erwin_12b}. As we will show below, in fact, ignoring disk breaks in decomposing galaxies has on average a significant effect on the measurement of disk scale lengths.  We also examine the effect of {\it not} including a break in disks that actually have a break on structural parameters such as the  bulge-to-total ratio (B/T), disk-to-total ratio (D/T), and bar-to-total ratio (Bar/T).  We also study how the radial position of the break varies along the Hubble sequence and as a function of B/T and galaxy mass to explore whether there is an universal physical mechanism that develops a break. To obtain a detailed census of the structural components and the properties of galaxies, we make use of {\it Spitzer} Survey of Stellar Structure (S$^4$G) \citep{sheth_10}. \s4g directly probes the old stellar population for over 2350 nearby galaxies using deep 3.6 and 4.5 {\micron} mid-infrared images taken with Infrared Array Camera (IRAC; \citealt{fazio_04}). This series of papers examines the structural properties of bars and disks in a subset of barred spiral galaxies from S$^4$G. 
One of the standard pipelines of \s4g (Pipeline 4, H. Salo et al. 2013, in prep) is decomposing all the galaxies into one to four sub-components using {\sc galfit} (\citealt{peng_02, peng_10}).  The purpose of our study, which also performs a 2D decomposition, is to use a different code, {\sc budda} (BUlge/disk Decomposition Analysis, \citealt{desouza_04, gadotti_08}), to add a disk break to the model fits, and to explore the relationship between the break and the co-evolution of the bar and inner disk.

The paper is organized as follows.  In the next section we describe our selection of a representative sub-sample of 144 barred galaxies from the \s4g for this study. In Section 3, we describe how 2D decompositions are produced. Section 4 presents the fits, as well as estimates of a number of structural parameters, and highlights interesting features from the decompositions. 
In Section 5, we present the results on the break radii, and show how they are associated with B/T, Hubble types, and 3.6 {\mus} magnitude and discuss our results.
In Section 6, we explore the effects of ignoring a break in the disk model for Type II (down-bending) disk galaxies. We quantify the differences in estimating the structural parameters when the break is properly accounted for and when the break is not considered in the model fit. 
We summarize our results in Section 7. The measurements obtained in this study will also be explored in forthcoming papers, where we will investigate the properties of bars and their host galaxies in the context of cosmological evolution.

\section{DATA AND SAMPLE SELECTION}
\setcounter{figure}{0}
\begin{figure}
\centering
\includegraphics[keepaspectratio=true,width=8cm,clip=true]{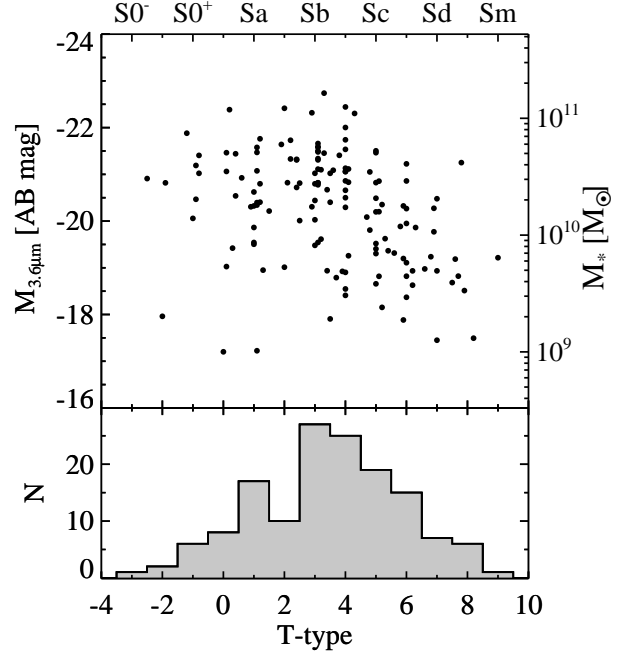}
\caption{Top: Hubble types, stellar masses and 3.6 \mus absolute magnitudes of the selected 144 galaxies. Bottom: Distribution of Hubble T types.}
\label{fig:ttype}
\end{figure}

We select our samples from \s4g \citep{sheth_10}. \s4g is a deep, volume-limited, magnitude-limited (total $B$-band magnitude$\leq$15.5) and size-limited (isophotal radius at 25 $B$-band mag per sq. arcsec larger than 1 arcmin) imaging survey of all galaxies with redshifts based on HI data satisfying these criteria at distances closer than 40 Mpc and galactic latitudes $|b|>30^\circ$, thus comprising over 2350 objects. Galaxies are mapped at 3.6 and 4.5 {\micron} and azimuthally averaged surface brightness profile typically reach a depth of 27 AB mag per square arcsec at 3.6 $\mu m$, corresponding to a stellar mass surface density of $\sim 1\,{\rm M}_\odot{\rm pc}^{-2}$. The images also extend large enough to cover 1.5$\times$D$_{25}$ for all galaxies. Hence these data are ideal  for a study of the structural properties of local galaxies.  

We chose galaxies that had been already processed by the first three \s4g  pipelines (Pipeline 1, 2, and 3, \citealt{sheth_10}) at the moment of this study (November 2011). In brief, Pipeline 1 (M. W. Regan et al. 2013, in prep) processes images and provides science-ready images. Pipeline 2 prepares mask images (to exclude foreground and background objects) for further analysis and Pipeline 3 derives surface brightness profiles and total magnitudes using IRAF ellipse fits (J. C. Mu{\~n}oz-Mateos et al. 2013, in prep).  Pipeline 4 (H. Salo et al. 2013, in prep) decomposes the 2D stellar distribution of galaxies into subcomponents with {\sc galfit} \citep{peng_02, peng_10}. 
To avoid the uncertainty caused by projection effects and/or disturbed morphologies, we excluded highly inclined ($b/a<0.5$), significantly disturbed, very faint, or irregular galaxies. Galaxies were also discarded if their images are unsuitable for decomposition due to contamination such as a bright foreground star or significant stray light from stars in the IRAC scattering zones. 
Then we chose barred galaxies from all Hubble types from S0 to Sdm using the numerical Hubble types from Hyperleda (\citealt{paturel_03}). The assessment of the presence of a bar was done visually by K.  Sheth, T. Kim, and B. de Swardt. Later, we also confirmed the presence of a bar by checking the MIR classification (\citealt{buta_10}, R. Buta et al. 2013, in prep), which are presented in Table 1. Except for UGC04393 that is classified as a peculiar galaxy by Buta et al., all of the selected galaxies are also classified as barred in the MIR classification. 93 galaxies are classified as SB (65\%), 31 galaxies are SAB (22\%), 5 galaxies are S$\rm \underline{A}B$ (3\%), 14 galaxies are SA$\rm \underline B$ (10\%), and one galaxy is classified as peculiar.
A total of 144 barred galaxies were selected which satisfy our criteria and we list our sample in Table 1 with basic information. 

In Figure~\ref{fig:ttype}, we plot Hubble types versus absolute 3.6 \mus magnitudes, and the distribution of Hubble T types. 
The stellar masses of the subsample vary from $10^9$ to $10^{11}$ $M_{\odot}$.  Stellar masses were derived using the 3.6 \mus magnitude according to Appendix A of \citet{munoz_mateos_13} that is based on the mass-to-light ratio from \citet{eskew_12}. While the sample does not cover the complete data set from \s4g, these selection procedures assure that the sample is {\bf (i)} representative of the local population of barred galaxies, and {\bf (ii)} suitable for structural analysis via image decomposition, meaning that the structural parameters can be accurately derived. We made use of 3.6 {\mus} images to derive structural parameters of galaxies. In this mid-infrared band the effects of dust are minimal and the data trace the bulk of the stellar mass distribution in the galaxies, with only a small local contamination (5--15\%) from AGB stars or hot dust surrounding red supergiants \citep{meidt_12a, meidt_12b}.

\section{DATA ANALYSIS: IMAGE DECOMPOSITION}
\subsection{Model fitting with {\sc budda}}
To produce the 2D galaxy fits in this study we make use of {\sc budda}. Details on the code and its usage can be found in \citet{desouza_04} and \citet{gadotti_08,gadotti_09}. Here we discuss the particular procedures adopted in this work to perform the decomposition. The images fed to the code for decomposition include a background contribution, necessary for a proper calculation of the model $\chi^2$. This can be critical to the measurements of structural parameters such as the bulge {\ser} index and the disk scale length, but it is also important for an accurate determination of the uncertainties. We measured sky background for each galaxy. In general, {\s4g} images have some amount of large-scale background fluctuations, and in a few cases deviations show up in the form of systematic spatial variations, such as gradients. In such cases, background gradients can be modeled using a 2D polynomial fit and can be removed (e.g., \citealt{comeron_11_4244}). However, in most cases the spatial variation of the large scale background deviation is random. Therefore, alternatively, the effect of large-scale background deviations can be estimated by performing model fits with various sky offsets (e.g., \citealt{busch_13xx}) and included in the error budget (e.g., \citealt{munoz_mateos_13}). In this study we did not take sky gradients into account in our model fits. However, we estimate uncertainties caused by the way the background was removed in the Section 3.2.
Avoiding foreground stars and background galaxies, we selected rectangular regions around the galaxies at around $R \sim 2 \times R_{25}$. Later we iteratively modified the chosen sky regions depending on the field of view of each image, and we took as background value the measured mode of the sky values in these regions.

Another critical parameter for {\sc budda} is the point spread function (PSF) full width at half maximum (FWHM).  {\sc budda} requires FWHM of PSF as an input parameter to create a Moffat function. We measured the PSF FWHM for each galaxy using a number of stars from each individual galaxy image. Average FWHM for our 144 galaxies is 1.76 arcsec (2.35 pixels in the \s4g image made with 0.75 arcsec/pixel), which translates to a physical scale of 180 pc at the median distance of our sample galaxies. Our decomposition can have up to four components: bulge, disk, bar and a nuclear point source, the latter is used to account for the possible presence of a nucleus (nuclear star cluster or non-stellar emission from an active galactic nucleus).

The bulge surface brightness profile is described by a S\'ersic function
\citep[][see \citealt{caon_93}]{sersic_63},

\begin{equation}
\mu_b(r)=\mu_e+c_n\left[\left(\frac{r}{r_e}\right)^{1/n}-1\right],
\end{equation}

\noindent where $r$ is the galactocentric distance, $r_e$ is the effective radius of the bulge, i.e., the radius that contains half of its
light, $\mu_e$ is the bulge effective surface brightness, i.e., the surface brightness at $r_e$, $n$ is the
S\'ersic index, defining the shape of the profile, and $c_n=2.5(0.868n-0.142)$.

In the case of disks without breaks, the disk surface brightness profile is described by a single exponential function:

\begin{equation}
\mu_d(r)=\mu_0 + 1.086r/h,
\label{eq:mu_nb}
\end{equation}

 \noindent where $\mu_0$ is the central surface brightness of disk, and $h$ is the characteristic disk scale-length. 
First, we examine surface brightness profiles of the galaxies to see whether there is a disk break by running {\sc ellipse}. If there is a change in the slope of surface brightness of disk, as in the case of Type II or Type III disks, an option to fit a second exponential profile is added in the model fit, as well as the break radius to {\sc budda}. In these cases, we labelled the disk as consisting of an inner disk and an outer disk, and the corresponding parameters are labelled as follows:

\begin{equation}
\begin{split}
\mu_d(r)= \begin{cases}
\mu_{\rm{0,in}} + 1.086r/h_{\rm in},  ~ \mbox {if } r \leq r_{\rm br} 
\\
\mu_{\rm{0,out}} + 1.086r/h_{\rm out}, ~ \mbox {if } r > r_{\rm br} 
\end{cases}
\label{eq:mu_wb}
\end{split}
\end{equation}

\noindent where $\mu_{\rm {0,in}}$ and $\mu_{\rm {0,out}}$ are the central surface brightness of inner disk and outer disk, respectively. $h_{\rm {in}}$ and $h_{\rm out}$ are the scale length of inner disk and outer disk, respectively. $r_{\rm br}$ is the break radius, also fitted as a free parameter by the code. 

The bar luminosity profile is also described by a S\'ersic function. For the bar,

\begin{equation}
\mu_{\rm Bar}(r)=\mu_{e,{\rm Bar}}+c_{n,{\rm Bar}}\left[\left(\frac{r}{r_{e,{\rm Bar}}}\right)^{1/n_{{\rm Bar}}}-1\right],
\end{equation}

\noindent where $c_{n,{\rm Bar}}=2.5(0.868n_{{\rm Bar}}-0.142)$, and the other parameters have definitions similar to those of the bulge above. Another bar parameter fitted by the code is the length of the bar semi-major axis, $r_{{\rm bar}}$, after which the bar light profile is simply truncated and drops to zero.

The nuclear point source is modeled as an unresolved point source convolved with the PSF profile. The FWHM of the point source profile has thus the same value as the PSF, and the only parameter fitted by the code is its peak intensity.

Bulge, disk and bar components are described by concentric, generalized ellipses
\citep[see][]{athanassoula_90}:

\begin{equation}
\left(\frac{|x|}{a}\right)^c+\left(\frac{|y|}{b}\right)^c=1,
\end{equation}

\noindent where $x$ and $y$ are the pixel coordinates of the ellipse points, $a$ and $b$ are the extent of its
semi-major and semi-minor axes, respectively, and $c$ is a shape parameter. Position angles and ellipticities
($\epsilon=1-b/a$) are fitted by the code for every component. When $c=2$ one has a simple ellipse, while when $c<2$ the ellipse is disky, and when $c>2$ the ellipse is boxy. For bulges and disks we fixed $c=2$ but
this parameter was left free to fit bars, since these components can be better described by boxy ellipses, in particular when the bar is strong \citep{athanassoula_90, gadotti_09}. $c$ is the boxiness parameter of the bar.

Some of our sample galaxies possess a nuclear bar or lens. There are 8 galaxies with a nuclear bar, and 10 galaxies with a nuclear lens according to the mid-Infrared classification from 3.6 micron images (\citealt{buta_10}, 2013 in prep, also see Table 1). However, due to the resolution (0.75 arcsec/pixel) and seeing limits (180 pc at the median distance), we may not be able to detect all known nuclear bars (see e.g., \citealt{erwin_02, comeron_10}). 
\citet{laurikainen_06} found that effects of ignoring the nuclear bar of disk galaxies are not significant with the Near-Infrared S0 Survey (\citealt{laurikainen_05, buta_06}) based on a higher resolution (0.23 arcsec/pixel) near-infrared images (see also \citealt{laurikainen_05,laurikainen_09}).
Also, with the current version of {\sc budda} one cannot not include two bars in the model fit. Therefore, we leave further analysis on nuclear bars for a future study.

To establish the code with initial guesses for the structural parameters we first performed ellipse fits using the {\sc ellipse} task in {\sc iraf}\footnote{{\sc iraf} is distributed by the National Optical Astronomy Observatories,
which are operated by the Association of Universities for Research in Astronomy, Inc., under
cooperative agreement with the National Science Foundation.}. Surface brightness profiles derived in such a manner helped us to visualize the initial estimates for a number of parameters, including the bulge effective radius, disk break radius, disk scale length and central surface brightness. Along with an analysis of the image, the profile is also used to estimate the end of the bar, i.e. the length of its semi-major axis \citep[see, e.g.,][]{athanassoula_02a, sheth_00, sheth_02, sheth_04, gadotti_07,menendez_delmestre_07}. 

We fit the bar with one component in this study. However, bars are known to exhibit two components -- an inner vertically thick part and an outer thin flat part \citep{athanassoula_05}. These two features can be seen on moderately inclined barred galaxies -- the thick part of the bar is boxy or peanut shaped and the outer thin part of the bar extends outwards into the disk and has ``spurs'' or handles (\citealt{athanassoula_06,laurikainen_07,erwin_13}). For the detailed shape of a bar, see the simulations (e.g., Figure 1 of \citealt{athanassoula_12_proc}). 
In some rare cases depending on the viewing angle, the inner boxy part of the bar and outer thinner part may appear to have different position angles (\citealt{athanassoula_06,erwin_13}).  In these instances it is not easy to fit the bar with just a single component (\citealt{erroz_ferrer_12}) because the position angle of the bar may change slowly with the radius and future models could attempt to fit these, but this is beyond the scope of the current study. We model the bar with just one component here.

For a barred galaxy where the bar is surrounded by a bright inner lens, we found that the code tried to fit the bar plus the lens instead of the bar alone. \citet{laurikainen_13} showed that the inner lenses are on average a factor 1.3 longer than bar and this behavior was also seen in the fits by {\sc budda}. There were also a few cases when the initial fit gave a larger bar length due to bright structures such as spiral arms, an inner ring, or a lens at the end of the bar. In these cases, we fixed the bar length to the value derived from our analysis of the ellipse fit profiles. 
Inspection of the ellipse profiles and the images was used to infer the presence of a bulge or a nuclear point source, which was then fed into the decomposition.  

If the initial guesses for the input values were overly far from those of the galaxy, we could obtain results that are not the best model fit, even though {\sc budda} returns a statistically best model fit with the minimum $\chi^2$ at each run. For all the galaxies we did not simply rely on the $\chi^2$ to judge the goodness of the fit. Multiple decompositions were performed for each galaxy and we compared models and images visually to make sure each component is well recovered, because there are some cases that fitted models do not represent well the galaxy. For example, even with the minimum $\chi^2$ model poor fits can be derived by either {\bf (i)} an overestimate of the bar length due to bright substructures at the bar ends, as discussed above, or {\bf (ii)} an overestimate of the bulge S\'ersic index due to a nuclear point source that was not included in the model, or {\bf (iii)} contamination in the bulge fit by a nuclear ring, or {\bf (iv)} in some cases when bars are not large enough, the fitted bulge model takes up the bar region and becomes highly elliptical, or {\bf (v)} when there is a bright lens or oval around the bar, {\sc budda} tries to account for the light from the lens or oval in the bar model and returns a too thick bar model, with low ellipticity, and finally, {\bf (vi)} difficulties in fitting the disk break (see more on this below).  
Therefore visual inspections on structural components were performed for every galaxy to ensure that the fitted model components do not overtake other components, and thus to assess which is the best fit.

The overestimate of the bulge {\ser} index was obvious when bulge models yielded $n>6$, while it was clear from the galaxy surface brightness profile that the bulge profile should have been closer to an exponential. This led to a bad fit in the outer parts of the bulge model and could be seen in the residual images.  We fixed this by adding a nuclear point source component to the model. (See \citealt{gadotti_08} for other instances of such problem.)  When a nuclear ring caused the bulge model to be incorrect we masked the nuclear ring from the input image. We show Hubble types and {\ser} indices of the bulge component of galaxies in Figure ~\ref{fig:sersic}. Our sample spans the full range of bulges in galaxies from  bulge-less to pseudo-bulges and classical bulges. 

\setcounter{figure}{1}
\begin{figure}
\centering
\includegraphics[keepaspectratio=true,width=8cm,clip=true]{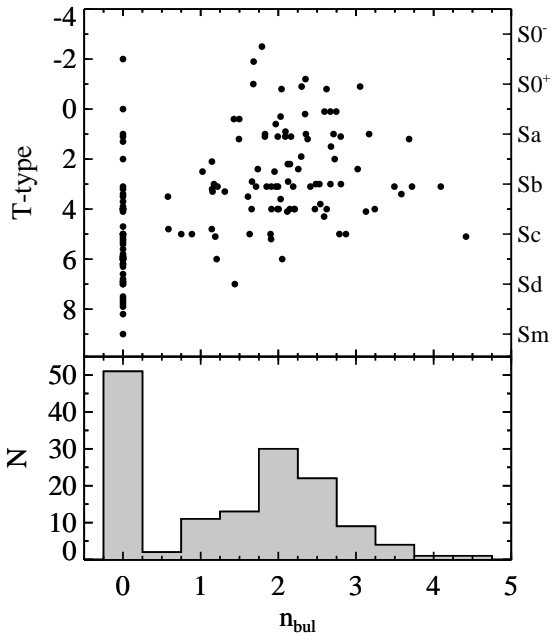}
\caption{Top: Hubble types and {\ser} indices of bulge components of the selected 144 galaxies. 
Bottom: Distribution of {\ser} indices of bulge components.}
\label{fig:sersic}
\end{figure}

In some cases the initial disk break radius was considerably far from the value expected from the galaxy surface brightness profile. The reason was that moderate changes in fitting the location of the break radius led to divergence in the $\chi^2$ minimization. In such cases we changed the algorithm to force slow variation in the iterations of the fitting that fixed the problem with the break radius.  

Some galaxies exhibit two breaks in their outer surface brightness profiles (e.g., \citealt{erwin_08, gutierrez_11, comeron_12, martin_navarro_12, munoz_mateos_13}). In these cases, because the latest {\sc budda} version can only fit one disk break for a galaxy, we decided to fit the inner break and ignore the outer break for two reasons: {\bf (i)} the minimization of $\chi^2$ is typically more substantial when the inner break is taken into account rather than the outer one and this gives a better model fit to obtain Bar/T and B/T. If we force {\sc budda} to fit outer break in the model, then it gives us poor Bar/T and B/T; and {\bf (ii)} because inner breaks seem to be related to the presence of the bar, in contrast to the outer break.

Different methods of estimating break and bar radii may give different results. In particular our results are based on 2D model fits, while most other studies on break are based on azimuthally averaged surface brightness fits. We compare our measurements of break and bar radii and those from \citet{munoz_mateos_13} in Appendix~\ref{ap:diff_method}. From this we conclude that the measurements of break and bar radii from these two studies agree within 20\% except for a few galaxies, and in such cases, the ratio of the two parameters ($r_{\rm br}/R_{\rm bar}$) may differ by up to 50\%.

\subsection{Uncertainties}
\setcounter{figure}{2}
\begin{figure}
\centering
\includegraphics[keepaspectratio=true,width=8cm,clip=true]{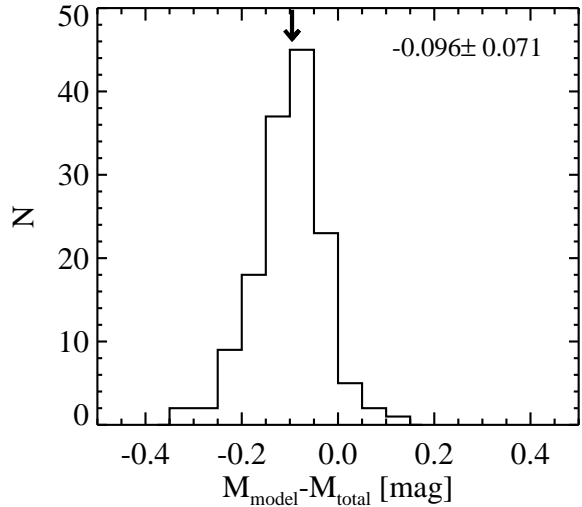}
\caption{Comparison between the total magnitudes of our models and the total magnitudes of the corresponding galaxies from \s4g Pipeline 3. The median and the standard deviation of the distribution are presented on the top right corner, and the arrow indicates the median of the distribution.
}
\label{fig:comp_model_galaxy}
\end{figure}
When galaxies are decomposed into multiple components the measurement of uncertainties for each structural component is non-trivial.   
We tested the impact of the sky background value on the model fits as follows:  we ran {\sc budda} fixing sky levels by adding and subtracting 1{\sig} of the sky background to the estimated sky level. 
The structural parameters that were sensitive to the sky background were the outer disk scale length, Bar/T, and B/T.
When the sky background was fixed to a 1{\sig} higher level, the outer disk scale length decreased by $\sim$10\% while the B/T and Bar/T increased by $\sim$8\%, and D/T decrease by $\sim$2\%.
In contrast, when the background was fixed to a 1{\sig} lower level, the outer disk scale length increased by $\sim$ 10\% while the B/T and Bar/T decreased by $\sim$8\% as D/T increase by $\sim$2\%. 
Other size-related parameters, such as break radius, bulge effective radius, and inner disk scale length vary less than 3\% due to a 1{\sig} change in the sky background value. These are thus the added errors in the error budget for each component only due to uncertainties in the estimation of the background value.

 {\sc budda} also gives statistical 1{\sig} uncertainty errors for each structural parameter.  
Errors are calculated after the code finds the global $\chi^2$ minimum. 
Successive variation is made on each parameter until the new $\chi^2$ reaches a threshold that is equivalent to 1{\sig} of a normal  $\chi^2$ probability distribution. 
In general, 1{\sig} uncertainty errors in our model fit range from 5 to 20\% and it
varies from component to component. Mean statistical 1{\sig} uncertainties of disk scale length and effective
radius of the bulge are 5 to 10\%. For break radius estimates,
mean 1{\sig} uncertainty is $\sim$ 17\%. Mean uncertainties of
position angles are 10 to 20\%, smaller for a bar and a disk
and larger for a bulge component. For ellipticities, it ranges from 5
to 15\%, smaller for bar and larger for bulge component.
Mean 1{\sig} uncertainty of bulge Sersic indices is $\sim$ 13\%, while
for bar {\ser} indices, uncertainty is $\sim$24\%. 
Since we fixed the bar length in our fits for some cases, the uncertainty in the bar length is not considered.

To evaluate our fits, we also have compared the total magnitudes of the models, calculated directly from the fit, and the total magnitudes of the galaxies, measured\footnote{Asymptotic magnitudes at 3.6 \mus, which can be downloaded from http://irsa.ipac.caltech.edu/data/SPITZER/S4G/} from the \s4g Pipeline 3. We plot this comparison in Figure~\ref{fig:comp_model_galaxy}. Models are slightly brighter by only $\sim$ 0.1 mag in the median. This may stem from the fact that there are often somewhat hollow regions in the disk surrounding the bar -- but inside the bar radius -- that are fainter than the model because stars that were in the disk around the bar are captured by the bar (e.g., NGC 4608 of \citealt{gadotti_08}), and the disk model does not account for that. Spiral arms or outer rings would compensate for that because we did not include models for such features separately. However, flux from those features are to some extent included in the disk models and this plot suggests that they are slightly less bright than necessary to compensate the hollow areas in the disk created by the bar.

\section{A CATALOG OF GALAXY STRUCTURAL PARAMETERS FROM 3.6 \mus IMAGES}
\setcounter{figure}{3}
\begin{figure*}
   \centering
  \vskip -0.3cm
  \includegraphics[keepaspectratio=true,scale=0.85,clip=true]{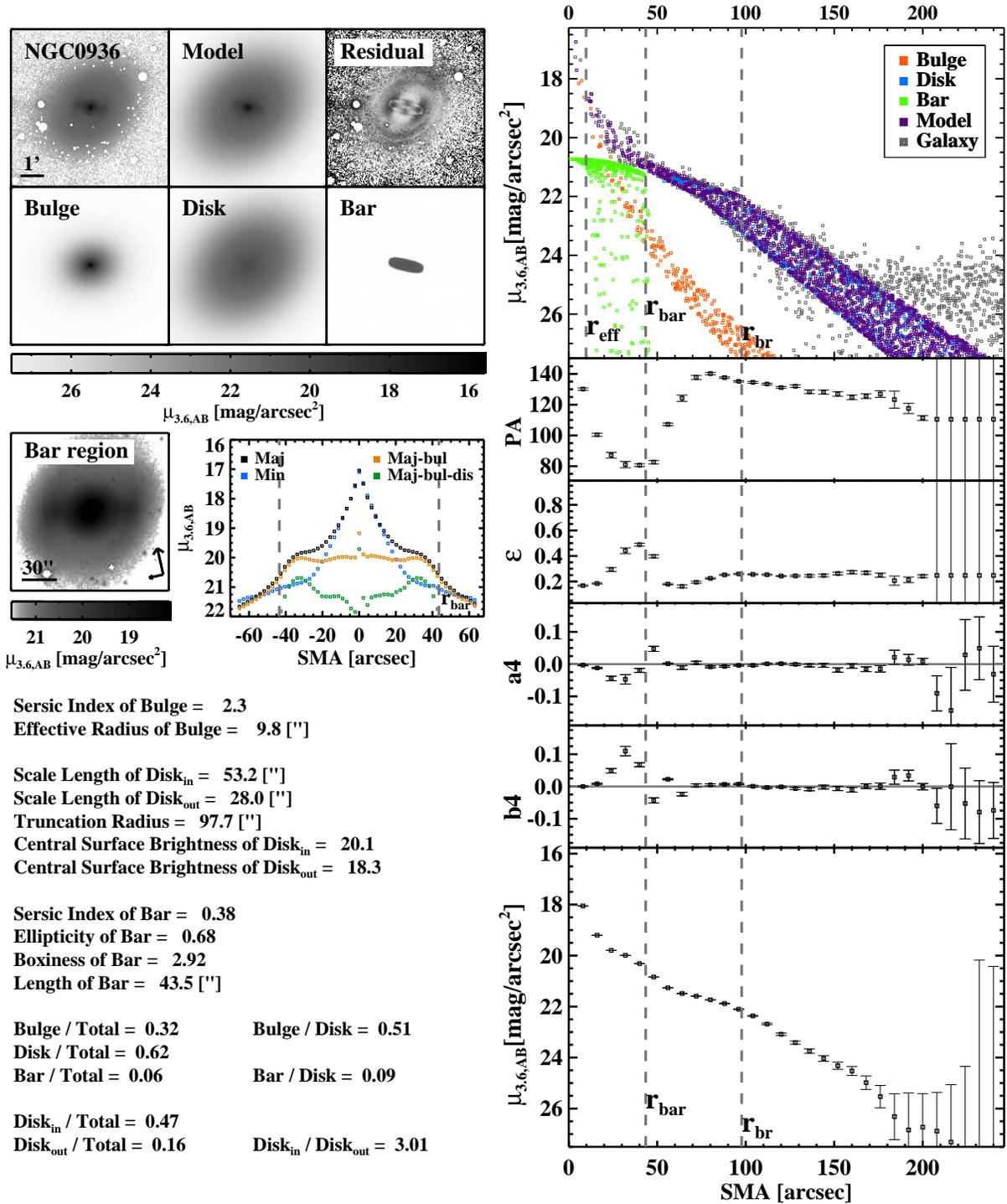}
   \vskip -1.4cm
   \caption{Results from the image decomposition of NGC 936. The top six panels on the left show, in thumbnail format, the original galaxy image, the total model image, a residual image obtained from subtracting the model image from the galaxy image, and separate images for bulge, disk and bar components only, as indicated. North is up, East to the left. All images, with the exception of the residual image, are shown in the same logarithmic scale. The display stretch of the residual image is chosen as to highlight major features. An extra panel with a zoom in the bar region is also shown. The next adjacent panel shows surface brightness profiles along the bar major and minor axes in the original image, and along the bar major axis when disk and bulge are subtracted. The vertical column of panels on the right shows radial profiles (from top to bottom): 2D surface brightness distribution of the galaxy and component models as indicated (every point is a pixel), and position angle, ellipticity, $a_4$, $b_4$ coefficient and azimuthally-averaged surface brightness from the original galaxy image, derived from ellipse fits. Vertical dashed lines mark the bulge effective radius, bar semi-major axis and break radius.
}
\label{fig:res}
\end{figure*}

Figure~\ref{fig:res} presents a summary page of the result for NGC 936 that has a bulge, and a bar, and show a disk break with a down-bending surface brightness profile. Thus the bulge, bar, and disk break are included in the model fit, and the result has different inner and outer disk scale lengths and an interpolated central surface brightness. From this figure, one can evaluate the resulting fit. 
A catalog of summary pages for all the other objects is presented in the Appendix~\ref{ap:summary}. In Figure~\ref{fig:res}, the first two rows of panels on the left show the original galaxy image at 3.6 {\micron}, the total model image, a residual image obtained from subtracting the model image from the galaxy image, and separate images of bulge, disk and bar model components. It is instructive to examine the model components separately and their appearance in the combined model image and compared to the original galaxy image. 

For some galaxies, the bulge model looks larger than in the galaxy image or in the total model, which is a sum of the bulge, disk, bar and central source.  This is because in the bulge model, the bulge is presented on the background that is set to a zero level, while the bulge in the total model and the galaxy image are on top of the disk and bar and thus the background around the bulge is higher due to the other components. Thus the bulge appears to be smaller in the total model and galaxy image. For illustrative examples of this see NGC 718 and NGC 1326 in Appendix~\ref{ap:summary}.
 
It is also instructive to search for sub-structures in the residual images -- these show all the structures not included in the model. Spiral arms thus appear conspicuous in many cases. It is important to stress that the residual images are produced with a {\em different} stretch compared to the images of other components to highlight the faint features and are fine-tuned for each galaxy. The difference in surface brightness between the model and residual substructures is typically 0.5 mag or less, consistent with the arguments presented in \citet{gadotti_08} that not accounting for the spiral arms in the decomposition is acceptable for the purposes of this study.

The residual images show a revealing substructure within many bars, in the form of a thinner dark stripe crossing the galaxy centre inside the bar (e.g., NGC 1433 and NGC 1452 in Appendix~\ref{ap:summary}). This feature was pointed out in \citet{gadotti_08} and plausibly represents an orbital family narrower than the main x1 family. We have not attempted to account for this feature but note that it suggests that more sophisticated bar models should include at least two components with different axial ratios. It also indicates that the Bar/T derived here may be systematically lower than the true ratio, though this difference is likely small. A wealth of detailed information is present in the residual images, including rings and substructure in spiral arms (see, e.g., NGC 4548 and NGC 5750).  Discussion of these substructures is beyond the scope of this paper.

In Figure~\ref{fig:res}, the third row of panels on the left shows a zoomed-in image of the bar region with the bar major axis rotated to be along the horizontal axis. In addition, we also show cuts along the bar major and minor axes, and along the bar major axis with the bulge (or disk) subtracted, as well as with the bulge and the disk subtracted (where applicable). A striking feature in these cuts is that some bars (e.g., NGC 936, NGC 1350, and NGC 5750) are better fitted with a flat surface brightness profile with S\'ersic index of about 0.5 or less. These bars indeed have a flat profile all the way to the central regions when the other components such as bulge and disk component are removed.  Conversely, some bars have steeper profiles, which resemble those of the disk (e.g., IC0167 and UGC04393). Structural parameters from the model fit are presented in the left bottom panel. 

The right-hand side of Figure~\ref{fig:res} shows a variety of radial profiles. The top one is a 2D surface brightness profile where every point is a pixel. This panel shows profiles for the galaxy, combined model and individual model components. This type of a plot has been used by \citet{laurikainen_05} and \citet{gadotti_08} as it displays virtually all the information in the image at the same time, as e.g. the different ellipticities of the different components, in contrast to profiles extracted from e.g. major-axis cuts. It also has the advantage that it enables us to distinguish between profile breaks that are caused from either lopsidedness or asymmetric spiral arms, and breaks in the profile that are indeed from the real disk break.
2D model fits are less hampered by such asymmetric features of galaxies in determining structural properties than one-dimensional (1D) profile fits, in particular for disk breaks. In our sample, we exclude strongly lopsided galaxies for our analysis by visual inspections, but there are still several galaxies that show some degree of lopsidedness, especially due to asymmetric spiral arm. Actually lopsidedness sometimes can create (pseudo) break in the radial surface brightness profile or change the location of the disk break when we examine azimuthally averaged 1D surface brightness profile. But these can be distinguished in 2D surface brightness profile by plotting all the pixels from the image.  In 2D surface brightness profiles, bright asymmetric spiral arms appear like stream lines which are brighter than surrounding interarm regions. (e.g., see right-top and bottom panel of Figure~13 in Appendix B for NGC 1637, ESO027-001). 2D model fits are less sensitive to the lopsidedness in determining break radii than 1D profile fits. 

Vertical dashed lines mark the bulge effective radius, bar semi-major axis and break radius, and are noted with $r_{\rm eff}$, $r_{\rm bar}$, and $r_{\rm br}$, respectively. The other radial profiles are derived from ellipse fits and correspond to position angle, ellipticity, Fourier coefficients a$_4$, b$_4$ and surface brightness, from top to bottom.

To measure systematic deviations of galaxy isophotes from perfect ellipses, Fourier analysis can be applied in general. 
High-order Fourier coefficients, such as a$_4$ and b$_4$, give us information about the shapes of the isophotes. These coefficients are obtained from the {\sc IRAF} task {\sc ellipse}, and a$_4$ is the coefficient that multiplies the term $\sin(4\theta)$ -- where $\theta$ is the ellipse eccentric angle, and these terms are used to describe the intensity distribution along the ellipse -- and b$_4$ is the coefficient that multiplies the term $\cos(4\theta)$. 
In particular, if b$_4$ is positive, isophotes are disky, whereas if b$_4$ is negative, the isophotes are boxy.  
a$_4$ indicates how much isophotes deviate from a bisymmetric structure, like e.g. in a parallelogram. a$_4$ indicates the presence of offset spurs in the bar region (\citealt{erwin_13}). If a$_4$ is positive, isophotes show counter-clockwise offset, while if a$_4$ is negative, isophotes show clockwise offset (for details, see \citealt{erwin_13}).

Table 2 presents the measures of structural parameters for all galaxies in the sample obtained from the 2D model fit with {\sc budda}, including bulge S\'ersic indices and effective radii, B/T ratios, bar lengths and ellipticities, Bar/T, disk scale lengths, break radii, etc. We have taken care to ensure the quality and uniformity of each of these measures through the combination of the {\it Spitzer} data products provided by \s4g and a careful inspection of the results.
In our sample, there are galaxies that possess an inner (pseudo)ring or lens or bright spiral arms, but also show a break farther out in the disk. 
We find that if we try to model these galaxies with a break, then {\sc budda} finds a break where the lens or inner ring ends, even though the real disk break is farther out. This is because a lens or an inner ring is much brighter than the break region and to minimize $\chi^2$ {\sc budda} identifies the edge of the lens or ring as a break. If we force {\sc budda} to find the real break, which is at a lower surface brightness level in the outer disk, then the code tries to add the lens or inner ring component to the bar, which is also inadequate. To better build the bar model rather than to add up a lens or inner ring to the bar model and to obtain better Bar/T and B/T, we do not force the code to find the real break radius. Therefore in these cases the derived break radius is actually the semi-major axis of the lens or inner ring and the derived scale length of inner (outer) disk for those galaxies correspond to the scale length of the disk inside (outside) the lens or inner ring. Note that these galaxies are mostly Type II.i disks, whose break occurs near or at the bar radius (\citealt{pohlen_06,erwin_08}, c.f., Type II.o disks have break well outside the bar).  These galaxies are marked with ``a'' in Table 2. We fit the inner break for such galaxies because they occur at higher (brighter) surface brightness level, and thus we can better estimate B/T, Bar/T, and D/T by fitting the inner break. 
This issue can be resolved if we are able to model fit a lens or inner ring component in the future with {\sc budda}. However, we choose to perform model fit with the current version of the code in this study. 
Finally, there are galaxies that display two breaks. 
If galaxies have a break at the inner ring or lens and show another break further out in the disk, we marked them with ``b''. If galaxies exhibit two breaks, but none of them are at the inner ring or lens, we marked such galaxies with ``c'' in Table 2. 
Galaxies listed with ``a'' and ``b'' are not taken into account for further analysis.
If we ignore the second break, we find 22 Type I, 120 Type II (49 Type II.i and 71 Type II.o), and 2 Type III disks.
 

\section{DISK BREAK RADII AND BAR RADII}
\subsection{Result}
\begin{figure}
\centering
\includegraphics[keepaspectratio=true,width=7.9cm,clip=true]{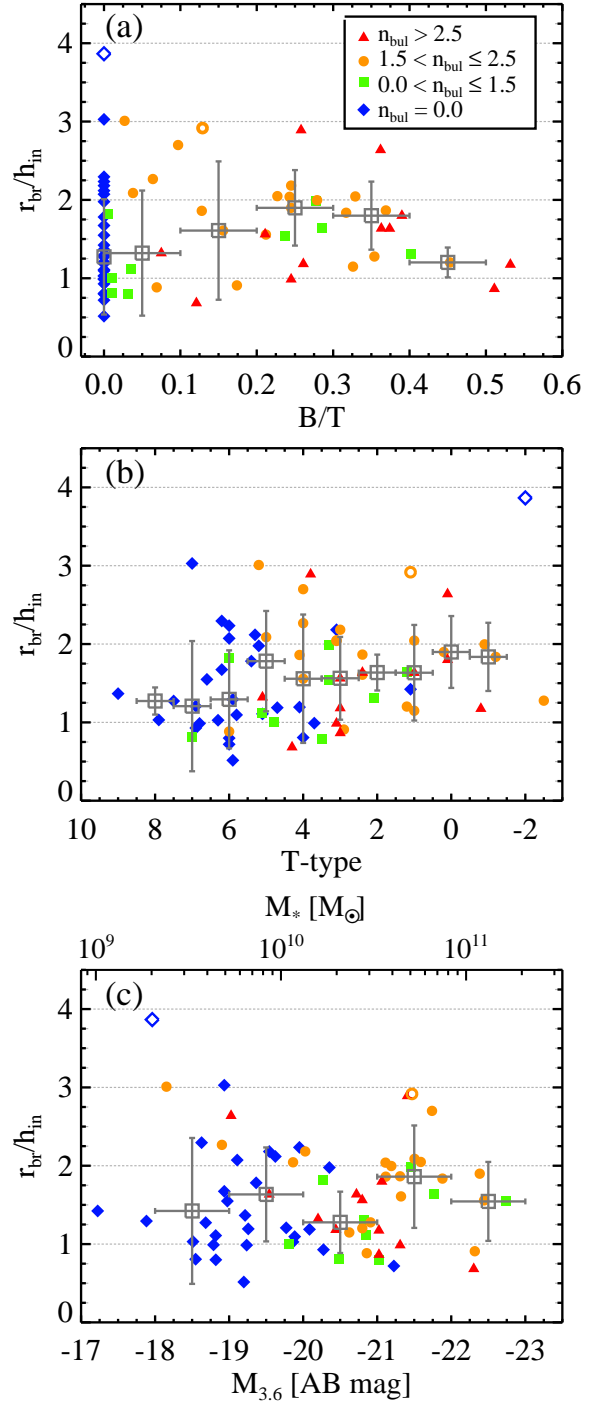}
\caption{Break radii scaled to disk scale length ($r_{\rm br}/h_{\rm in}$) for Type II (down-bending) and Type III (up-bending) disk galaxies as a function of (a) bulge-to-total, (b) Hubble type, and  (c) stellar mass estimated using 3.6 \mus magnitudes. {\ser} indices of bulge components are color-coded. Type II disk galaxies are in filled symbols, and Type III disk galaxies are in open symbols. Galaxies in which we identified only one break in each galaxy are presented. Grey squares indicate medians at each bin that is covered with the horizontal error bar, while vertical error bar spans the standard deviation of break radius in units of disk scale length at each bin.}
\label{fig:rbr_hi}
\end{figure}

We examine how the break radius ($r_{\rm br}$) changes with B/T, Hubble types, and 3.6 {\micron} magnitude of galaxies. $r_{\rm br}$ is the radius where the slope of surface brightness of disk changes (Eq.~\ref{eq:mu_wb}). $r_{\rm br}$ is associated with galaxy stellar mass (\citealt{comeron_12, munoz_mateos_13}) in the sense that more massive disks exhibit breaks at larger radii than less massive disks. However, when normalizing a break radius by a bar radius ($R_{\rm bar}$), \citet{munoz_mateos_13} find that the trend with mass disappears. Instead the range of possible break radius to bar radii ($r_{\rm br}/R_{\rm bar}$) is strongly dependent on the total stellar mass. $r_{\rm br}/R_{\rm bar}$ of massive galaxies ($\ge 10^{10} M_{\sun}$) spans a range from 2--3, whereas in less massive galaxies  $r_{\rm br}/R_{\rm bar}$ spans a range from 2--10. However, it has not been explored how $r_{\rm br}$ changes as a function of B/T.

\begin{figure*}
\centering
\includegraphics[keepaspectratio=true,width=17cm,clip=true]{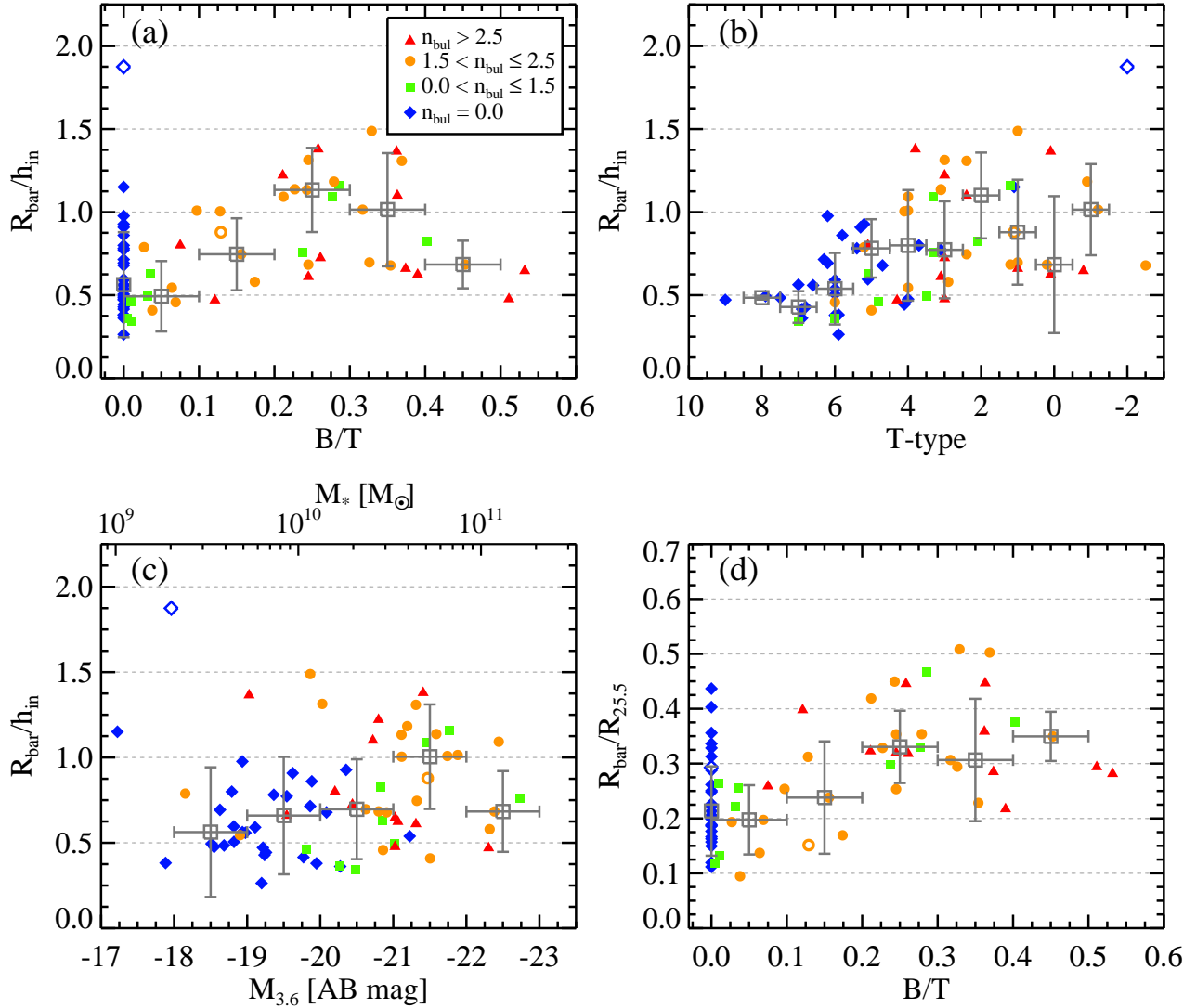}
\caption{
Deprojected and normalized bar radii for Type II (down-bending) and Type III (up-bending) disk galaxies. 
(a) Deprojected bar radii scaled to inner disk scale length ($R_{\rm bar}/h_{\rm in}$) as a function of bulge-to-total, (b) $R_{\rm bar}/h_{\rm in}$ as a function of Hubble type, (c) $R_{\rm bar}/h_{\rm in}$ as a function of stellar mass from 3.6 \mus magnitudes, and (d) deprojected bar radii scaled to the semi-major axis at $\mu_{3.6}$ = 25.5 AB mag arcsec$^{-2}$ ($R_{\rm bar}/R_{\rm 25.5}$) as a function of bulge-to-total. Points are color-coded with {\ser} indices.
Bar radii are deprojected ($R_{\rm bar}$) analytically according to \citet{gadotti_07}. Type II disk galaxies are in filled symbols, and Type III disk galaxies are in open symbols. Galaxies that we identified only one break in each galaxy are plotted. Grey squares indicate medians at each bin that is covered with the horizontal error bar, while vertical error bar spans the standard deviation of bar radius in units of $h_{\rm in}$ or $R_{\rm 25.5}$ at each bin.
 }
\label{fig:rbar_hi}
\end{figure*}

\begin{figure*}
\centering
\includegraphics[keepaspectratio=true,width=17cm,clip=true]{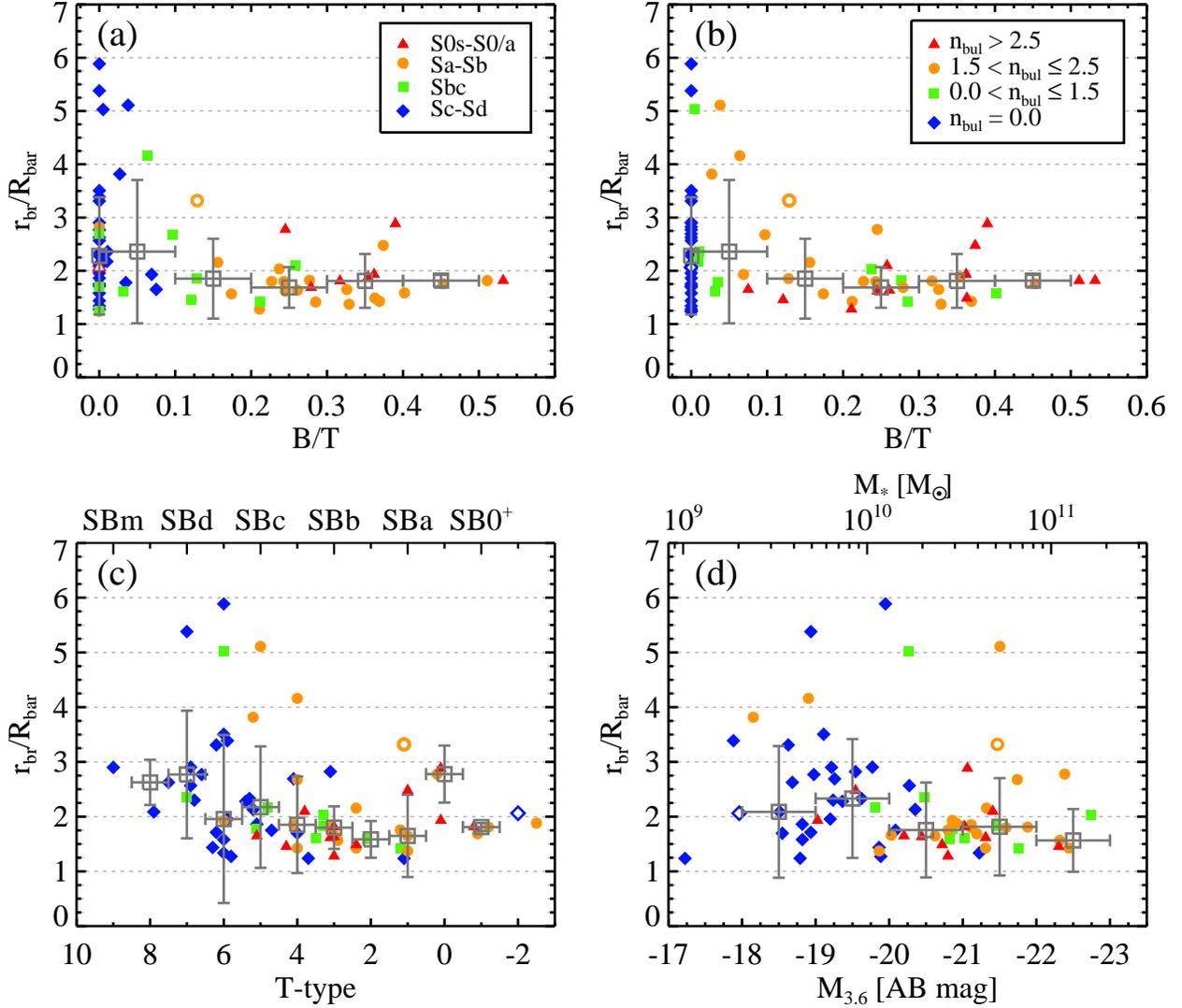}
\caption{Break radii scaled to bar radii ($r_{\rm br}/R_{\rm bar}$) for Type II (down-bending) and Type III (up-bending) disk galaxies as a function of bulge-to-total (a) and (b), Hubble type (c), and stellar mass (d) estimated using 3.6 \mus magnitudes. Hubble types are color-coded in (a), while {\ser} indices of bulge components are color-coded in (b),(c), and (d). Type II disk galaxies are in filled symbols, and Type III disk galaxies are in open symbols. 
Galaxies in which we identified only one break are presented in this figure and galaxies that have more than one break are excluded. Bar radii are deprojected ($R_{\rm bar}$) analytically according to \citet{gadotti_07}. Grey squares indicate medians at each bin that is covered with the horizontal error bar, while vertical error bar spans the standard deviation of break radius in units of $R_{\rm bar}$ at each bin. 
}
\label{fig:rbr}
\end{figure*}

We examine the break radii scaled to inner disk scale length ($r_{\rm br}/h_{\rm in}$) and plot them in Figure~\ref{fig:rbr_hi} as a function of B/T as well as Hubble type, and stellar mass of the galaxy (calculated using the absolute magnitude at 3.6 {\micron}). Points are color coded by {\ser} indices of bulge components. We show the median $r_{\rm br}/h_{\rm in}$ with grey squares at each bin and the standard deviation of the normalized break radius at each bin with vertical error bars. Type II disks are shown with a filled symbol and Type III with an open symbol.
In our Figures~\ref{fig:rbar_hi} and \ref{fig:rbr} we only plot galaxies identified with single disk break and do not plot those with two breaks or where a break is caused by an inner ring or a lens. While there is a significant scatter, the median of $r_{\rm br}/h_{\rm in}$ at each bin shows only a small increase with B/T at 0$<$B/T$<$0.5, and no trend with mass. 

We deprojected the bar lengths analytically following \cite{gadotti_07}, assuming that bars can be represented by an ellipse\footnote{In reality, we find that the bar shapes are predominantly boxy but this does not affect the derivation of the deprojected bar length. Details on the shapes of bars will be discussed in a subsequent paper (T. Kim et al. 2014, in prep.)}. The deprojected bar lengths are expressed as $R_{\rm bar}$. We show deprojected bar radii scaled to the inner disk scale length ($R_{\rm bar}/h_{\rm in}$) and deprojected bar radii normalized by the semi-major axis at $\mu_{3.6}$ = 25.5 AB mag arcsec$^{-2}$ ($R_{\rm bar}/R_{\rm 25.5}$) in Figure~\ref{fig:rbar_hi}. Our data show that the $R_{\rm bar}/h_{\rm in}$ and $R_{\rm bar}/R_{\rm 25.5}$ increases with B/T, from later to earlier type galaxies. However no trend is obvious with galaxy luminosity/mass.  

In Figure~\ref{fig:rbr}, we plot break radii normalized by the deprojected bar radius ($r_{\rm br}/R_{\rm bar}$).
Figure~\ref{fig:rbr}(a) is color coded by Hubble types while other panels, Figure~\ref{fig:rbr}(b), (c), and (d), are color coded by {\ser} indices of bulge components. 
Galaxies with B/T$>$0.1 span $r_{\rm br}/R_{\rm bar}$ $\sim$ 1--3 with most around $\sim$ 1.8.  The $r_{\rm br}/R_{\rm bar}$ does not vary much as a function of B/T. However, galaxies with small bulges (B/T$<$0.1) have a $r_{\rm br}/R_{\rm bar}$ that ranges from 1 to 6. There are also a number of galaxies that have a large $r_{\rm br}/R_{\rm bar}$ and this leads lower B/T galaxies to have larger median $r_{\rm br}/R_{\rm bar}$ than galaxies with B/T$>$0.1.
Late Hubble types (T$>$3), i.e., less massive galaxies exhibit a wider scatter than early types (T$\leq$3). Breaks in some of later type disks (less massive galaxies) occur farther out than in earlier types (more massive galaxies) and show a wide distribution in $r_{\rm br}/R_{\rm bar}$. Galaxies with higher bulge {\ser} indices (classical bulge) have $r_{\rm br}/R_{\rm bar}$ $\sim$ 1--3. In fact, $r_{\rm br}/R_{\rm bar}>$3 galaxies have bulge {\ser} indices less than 2.5. However, there is no clear trend on normalized break radii with bulge {\ser} indices on the whole. It is interesting that $r_{\rm br}/R_{\rm bar}$ of high B/T galaxies (B/T$>$0.1) form tighter sequence as a function of B/T than the cases shown along the Hubble types or stellar mass. Moreover, $r_{\rm br}/R_{\rm bar}$ of higher B/T galaxies do not vary much in $0.1<$B/T$<0.5$.
Galaxies that have $r_{\rm br}/R_{\rm bar}>$5 are NGC 1232, NGC 5584, NGC 5669, and PGC003853, all of them are late types.

In Appendix~\ref{ap:outering}, we examine where disk breaks occur. By comparing break radii with outer ring radii, we find that for more than half of barred galaxies with outer ring, disk breaks occur at the outer ring. Because an outer ring is thought to be at the outer Lindblad Resonance (OLR) of the bar (\citealt{schwarz_81, buta_91, buta_95}), this implies that disk breaks arise at the OLR of the bar for those galaxies.
 

\subsection{Discussion}
Figure~\ref{fig:rbr} shows that higher B/T galaxies form a tighter
sequence of $r_{\rm br}/R_{\rm bar}$, whose mean values are
roughly constant. At low B/T, while some galaxies exhibit $r_{\rm
  br}/R_{\rm bar} \sim 2$, there are also galaxies with a large
$r_{\rm br}/R_{\rm bar}$ compared to higher B/T galaxies, in agreement
with other studies that explored break radii along the Hubble sequence
or as a function of stellar mass (e.g., \citealt{pohlen_06, erwin_08,
  gutierrez_11, munoz_mateos_13}). This implies that, for low B/T
galaxies, there may be another mechanism to drive the break.
Recently, \citet{munoz_mateos_13} found that $r_{\rm br}/R_{\rm bar}$
shows a bimodal distribution, showing peaks at $\sim$2 and 3.5 in
galaxies more massive than $10^{10}${\Msun}.  They also showed that
breaks can be found at large radii, $r_{\rm br}/R_{\rm bar}$ up to
10. They argued that the first peak at 2 is likely associated with the
bar OLR, whereas the second peak at 3.5 is from a coupling of
resonances between the spiral arm and bar pattern speed. Hence in their
interpretation, \citet{munoz_mateos_13} could explain the distribution
of breaks without invoking a star formation threshold or
efficiency. It thus becomes clear that we need to understand better
the ratio of characteristic lengths, e.g., the bar length, the radius
at which the break occurs and the various relevant radii 
such as the corotation (CR) and the OLR. These obey specific rules set by
dynamics, contrary to break radii that are set by star formation. We
will thus first summarize some previous  results on these lengths, 
before discussing more specifically our results.      

By assuming a mathematically simple rotation
curve, \citet{athanassoula_82} found that the possible
range of resonance radii ratios (e.g., $R_{\rm OLR}/R_{\rm CR}$) depends
on the shape of the rotation curve, and thus on galaxy type. 
This was used by \citet{munoz_mateos_13},
who made use of 3.6 \mus photometry to estimate the radius where the rotation
curve of the galaxy reaches the flat regime ($r_{\rm flat}$) and showed
that $r_{\rm flat}$ is larger for late types, low-mass systems and thus the OLR
may be located farther away from corotation compared to early types. 
This argues that in late type galaxies $r_{\rm br}/R_{\rm bar}$ 
can reach larger values than in early type galaxies. 

Observational studies have shown that bars become longer
as they evolve (\citealt{elmegreen_07}), as predicted by simulations.
\cite{athanassoula_03} showed that as a bar looses angular momentum
to the outer disk and halo resonances, it will get longer and
stronger. Its pattern speed will decrease (e.g., \citealt{little_91a, debattista_00, athanassoula_03, martinez_valpuesta_06}).
Note, however, that the presence of gas may slow down this process 
(e.g., \citealt{berentzen_07, athanassoula_13a}; E. Athanassoula et al. 2013b, in prep).
As the bar grows, the position of the CR is pushed outwards to make
space for the newly trapped orbits in its outer parts. The OLR also
moves outward and thus the break radius, if it is indeed linked to the
bar OLR, will also move outwards so that no important changes in 
$r_{\rm br}/R_{\rm bar}$ can be expected that way. 

In general it is expected that the corotation radius ($R_{\rm CR}$) should be in the range $R_{\rm bar}$ -- $1.4 \times R_{\rm bar}$. 
However, some observational and theoretical works have voiced the possibility
that bars in late type galaxies may be shorter compared to the main
resonant radii than those in early type disks \citep{elmegreen_85, combes_93, rautiainen_05, rautiainen_08}. 
Since late type disks have a low B/T, this will imply that they will have increased
values of $r_{\rm br}/R_{\rm bar}$ and thus could provide part of the
explanation of the difference between the galaxies with B/T>0.1 and
those with B/T<0.1.  

Our results give rise to some further interesting implications and
speculations. Figure~\ref{fig:rbar_hi}(d) shows clearly that the
bar length (normalized by $R_{\rm 25.5}$) is an increasing function of B/T. 
This is in good agreement with previous observational results 
(e.g., \citealt{elmegreen_85, laurikainen_07, menendez_delmestre_07, gadotti_11, cheung_13xx}) 
as well as with results of simulations. 
It is important to note that Figure~\ref{fig:rbar_hi} includes both classical and disky pseudo-bulges, so our interpretation has to include both types of bulges.  

Let us first consider classical bulges. \citet{athanassoula_02a} compared two models identical in everything except for the
existence/absence of a classical bulge component and found that the
bar is considerably longer and stronger in the model with a
bulge. This was confirmed and explained by \citet{athanassoula_03}, as due
to the extra angular momentum that can be absorbed by the bulge
component, which leads to a considerable increase of the angular
momentum emitted by the bar and thus a considerable increase of the
bar length and strength.

The interpretation for disky pseudo-bulges is more straightforward.
Let us recall that longer bars are expected to be also stronger
(see \citealt{athanassoula_12_canary} for a review). Furthermore, longer and 
stronger bars will push more gas inwards to the central regions (\citealt{athanassoula_92}), and will thus form stronger disky pseudo-bulges. 

Thus simulations predict, both for classical and for
disky pseudo-bulges, the existence of a correlation between
the bar length and the bulge mass. This is indeed what we see in our
data. What simulations still need to explain, however, is why the two types of
bulges lie on the {\it same} correlation.  

Let us now turn to Figure~\ref{fig:rbar_hi}(a). Here
we normalize the bar length by the inner disk scale length ($R_{\rm bar}/h_{\rm in}$) and we find that for the heaviest of bulges there is a
decrease of the bar length, as found also by \citet{cheung_13xx}. 
This could be explained by the fact that $h_{\rm in}$ is also 
evolving with time and depends on the angular momentum exchange. 
Comparing Figure~\ref{fig:rbar_hi}(b) and Figure~\ref{fig:rbar_hi}(c)
could thus suggest that the increase of the classical bulge mass
beyond a certain limit influences $h_{in}$ more than the bar
length. Alternatively, the increase of the bar length could be limited
by the extra concentration of the heavy-mass classical bulges. 
More work is necessary to confirm or reject these possibilities.

Figure~\ref{fig:rbr} shows that for B/T> 0.1 the mean and dispersion of the
$r_{\rm br}/R_{\rm bar}$ values have no clear dependence on the relative bulge
mass. In light of what  
we discussed above about the dependence of the bar length on the
relative bulge mass, this should imply that $r_{br}$ and $R_{bar}$ have a
similar dependence on the bulge mass. This is reasonable if the break
is linked to the bar, for example if it occurs at the bar OLR and the
bar is linked to CR. Our results thus give further corroboration to
this possibility.

For B/T<0.1, we find that both the mean and dispersion of the
$r_{\rm br}/R_{\rm bar}$ are larger than those of larger B/T galaxies.
Thus it is reasonable to assume that the bar is not the main
driving agent here. This is further corroborated by the fact that a
number of our galaxies have $r_{\rm br}/R_{\rm bar}$ larger than 3 or 4, while
a few reach values of 5 or 6. At such distances from the bar, its force is considerably diminished, 
so that it can not be assumed to the
driving force for the formation of any structure. Even spirals coupled
to the bars by their resonances (\citealt{tagger_87, sygnet_88, rautiainen_99, quillen_11, minchev_12a})
will have difficulty reaching such distances, unless there is a set 
of several such coupled spirals.     

Our results are thus in good general agreement
with previous observational and theoretical results while adding new
information on the formation of disk breaks. Nevertheless,
this evolutionary picture, appealing though it may be, leaves many
questions open for further study.

We therefore conclude that breaks in galaxies with B/T $\geq$ 0.1 are
bar-driven. The break to bar radius remains rather constant once a
prominent bulge has formed and evolves to B/T$\sim$0.5. 
In galaxies with inconspicuous bulges or bulgeless galaxies, there are
both high and low $r_{\rm br}$/$R_{\rm bar}$, implying that the breaks
may be due to the bar-only resonance, to a spiral-bar coupling, and/or
to star formation thresholds. The latter mechanisms allow larger
$r_{\rm br}/R_{\rm bar}$ ratio and increase the scatter of the
distribution of $r_{\rm br}/R_{\rm bar}$. 


\section{CHANGES IN THE STRUCTURAL PARAMETERS DUE TO DISK BREAK}

\subsection{Result}
\begin{figure}
\centering
\includegraphics[keepaspectratio=true,width=8cm,clip=true]{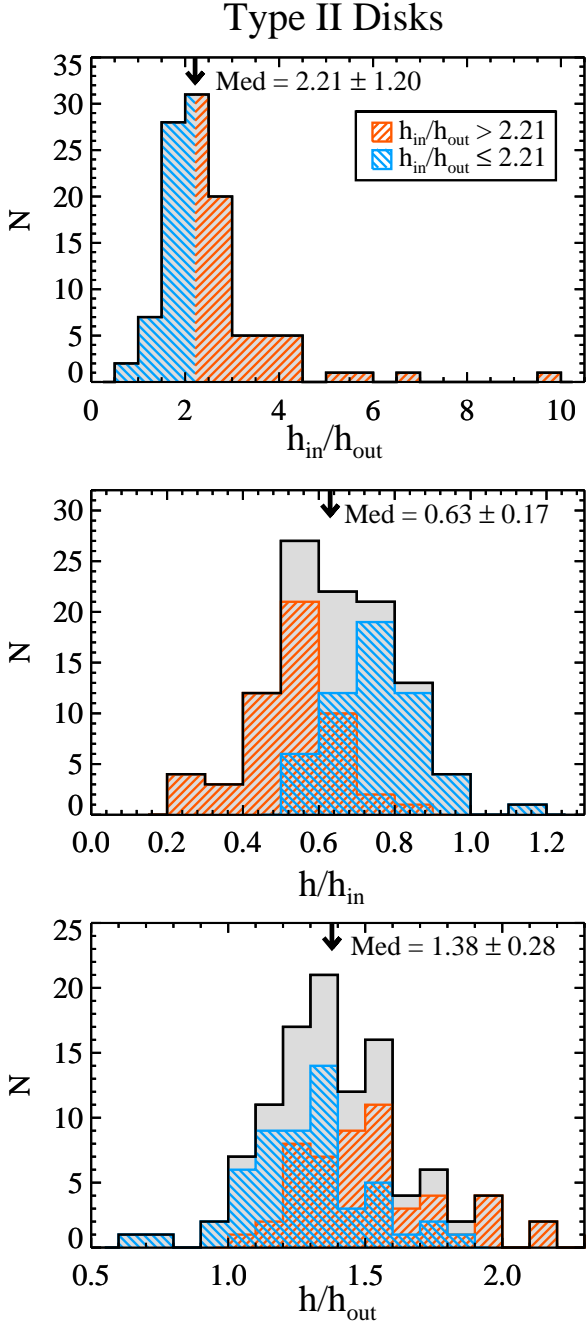}
\caption{Comparisons of disk scale length from the model fit that ignored a break and included the disk break properly in the model fit. Distributions of disk scale length ratios for Type II disk galaxies are shown.
Top panel: Inner disk scale length to outer disk scale length  ($h_{\rm in}/h_{\rm out}$).
Middle panel: disk scale length from the fit that ignored the break ($h$) to the inner disk scale length ($h_{\rm in}$).
Bottom panel: disk scale length from the fit that neglected the break ($h$) to the outer disk scale length ($h_{\rm out}$). Galaxies that have stronger breaks ($h_{\rm in}/h_{\rm out} >$ 2.21) are plotted in red, and galaxies that have weaker breaks ($h_{\rm in}/h_{\rm out}$ $\leq$ 2.21) are in blue. In each panel, an arrow indicates the median of the distribution, and the standard deviation is also presented.
}
\label{fig:comp_rscl}
\end{figure}

The majority of disk galaxies exhibit either down-bending (Type II) or up-bending (Type III) disk profiles \citep[]{pohlen_06, hunter_06, erwin_08, gutierrez_11, maltby_12a}. Thus it is crucial to account for the break in the disk model.
Modeling a galaxy ignoring a break is similar to modeling a Type II galaxies with Type I profile.
However, Type I and Type II disk galaxies show different structural properties as shown in Figure 5 of \citet{munoz_mateos_13}.
To understand how inner and outer disks form and evolve, it is necessary to derive structural parameters of these two disks suitably.
Accounting for disk breaks in 2D fits may lead to only small changes in some of the derived parameters. \citet{erwin_12b} have shown that for the particular case of NGC 7418 including a break in the disk changes the B/T from 0.016 to 0.017 only. 
However, in this section we demonstrate that the effect on disk parameters when not accounting for disk breaks can be substantial, especially for Type II down-bending disks. To evaluate the importance of including disk breaks in 2D model fits, we produce fits with and without a break for Type II disk galaxies.

Before assessing the effect of ignoring a break in model fitting on the measurement of disk scale lengths and luminosity ratio of the bar and bulge to the disk, we evaluate how much the inner disk scale length ($h_{\rm in}$) and outer disk scale length ($h_{\rm out}$) differ. To check the strength of the break ($h_{\rm in}/h_{\rm out}$), we show the distribution of $h_{\rm in}/h_{\rm out}$ for Type II (down-bending) disk galaxies in the top panel of Figure~\ref{fig:comp_rscl}. 
We find that the median inner disk scale length is $\sim$ 2.21 times larger than the median outer disk scale length, with the standard deviation of 1.20. This is in agreement with the result of \citet{pohlen_06} who found $h_{\rm in}/h_{\rm out} \sim$ 2.1 $\pm$ 0.5 using SDSS $g$\'{}- and r\'{}- band image for late type disk galaxies (Sb -- Sdm). \citet{martin_navarro_12} calculate the stellar surface mass density profile and they find ${h_{\rm in}/h_{\rm out}}_{log {\Sigma}} \sim$1.6 for edge-on galaxies. 

We compare the disk scale lengths that were obtained from a model fit ignoring the break $(h)$ and the disk scale lengths from the model fit that included the break, i.e $h_{\rm in}$ and $h_{\rm out}$. We show the distribution of $h/h_{\rm in}$ and $h/h_{\rm out}$ in the middle and bottom panels of Figure~\ref{fig:comp_rscl}, respectively. We find that $h/h_{\rm in}$ varies from 0.3 to 1.0. If we ignore the break in the model fit the recovered disk scale length is smaller than the inner disk scale length and larger than the outer one. 
The median recovered disk scale length is 0.63 times the inner disk scale length and 1.38 times the outer disk scale length with a standard deviation of 0.17 and 0.28 respectively.

The differences of the disk scale length from a fit that ignored the break to inner and to the outer disk scale length ($h/h_{\rm in}$ and $h/h_{\rm out}$) depend on the strength of the break. We divide galaxies into two groups: galaxies with $h_{\rm in}/h_{\rm out} > 2.21$ (median of $h_{\rm in}/h_{\rm out}$ for Type II samples) and $h_{\rm in}/h_{\rm out} \leq 2.21$. We overplot the galaxies with stronger break, i.e. the inner disk scale length is much longer than the outer one, in red and galaxies with weaker break in blue in Figure~\ref{fig:comp_rscl} to examine how much different $h/h_{\rm in}$ and $h/h_{\rm out}$ vary for those two groups. If the break is strong, then the difference between the $h$ and $h_{\rm in}$ becomes more pronounced. This also holds for $h$ and $h_{\rm out}$, as can be seen in Figure~\ref{fig:comp_rscl} with the corresponding histograms plotted.

\begin{figure}
\centering
\includegraphics[keepaspectratio=true,width=8.4cm,clip=true]{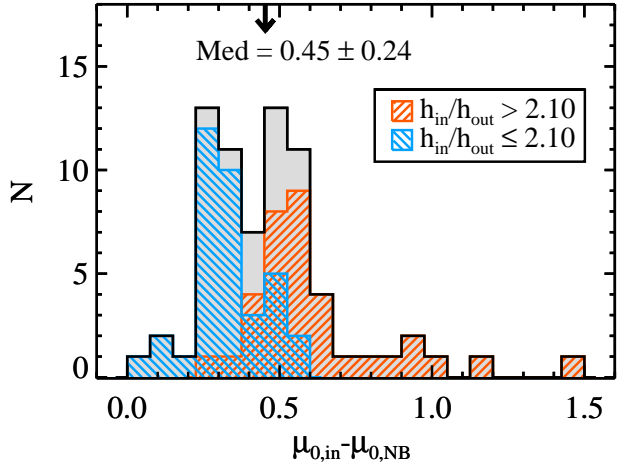}
\caption{Comparisons of central surface brightness of the inner disk ($\mu_{0,\rm in}$) from the model fit that included the disk break properly and those from the model fit with no break ($\mu_{0,\rm NB}$) for Type II disk galaxies. Central surface brightness of disks is corrected for inclination.
 }
\label{fig:comp_rscl9}
\end{figure}

Central surface brightness of disks (Eq.~\ref{eq:mu_nb} and ~\ref{eq:mu_wb}) also changes if we do not account for a break.
We examine how much difference it will make to ignore a break in estimating central surface brightness of inner disk and plot them in Figure~\ref{fig:comp_rscl9}. Central surface brightness of disks is corrected for inclination, $\mu_{\rm cor}= \mu_{\rm obs} - 2.5 log (b/a)$, as in \cite{munoz_mateos_13, sorce_13}. We find that if we introduce a break in the disk model fit, the median central surface brightness of the inner disk becomes $\sim$0.45 mag fainter in median than that of the disk modeled without a break. 

\begin{figure}
\centering
\includegraphics[keepaspectratio=true,width=8.4cm,clip=true]{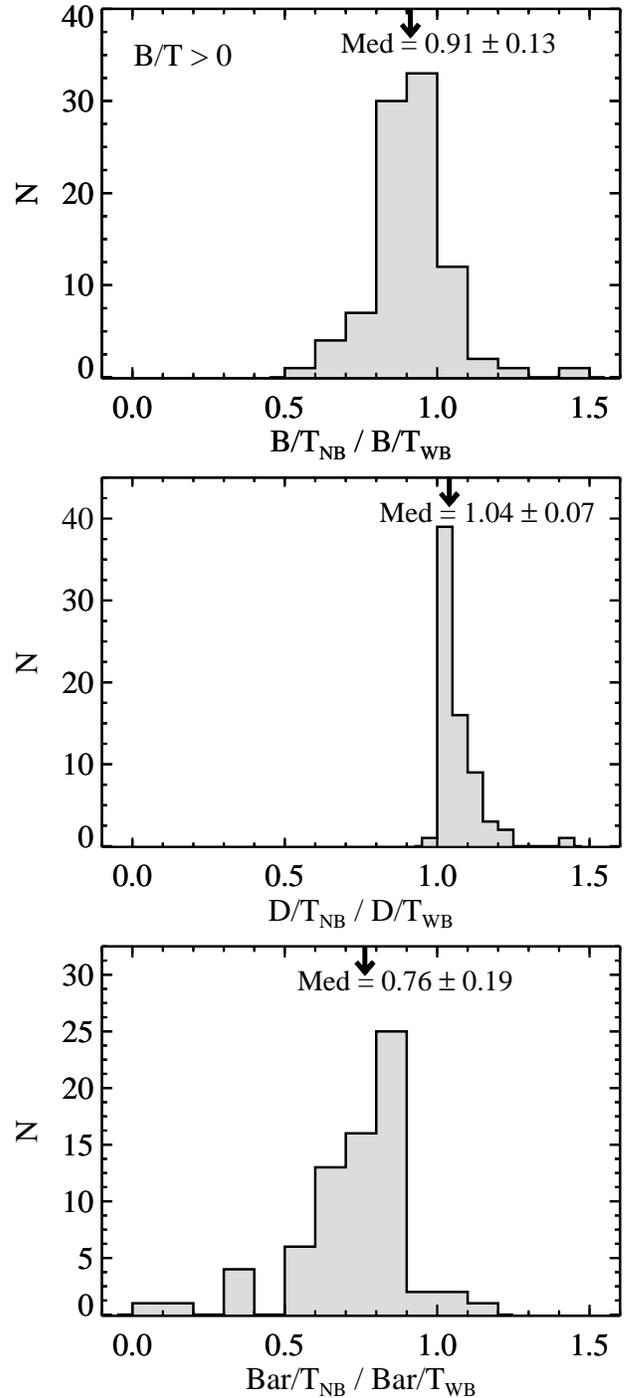}
\caption{Comparisons of galaxy component luminosity ratios for Type II disk galaxies when i) galaxies are modeled with a disk model that includes a break and ii) galaxies are modeled without a break, i.e. just with a single exponential function.
Top panel: Bulge-to-total (B/T) estimated from a disk model with a single exponential function, i.e., with no break $(B/T_{\rm NB})$ over bulge-to-total luminosity ratio obtained from a disk model with the disk break $(B/T_{\rm WB})$.
Middle panel: same as top panel, but for disk-to-total (D/T).
Bottom panel: same as the top panel, but for bar-to-total (Bar/T).}
\label{fig:comp_rscl5}
\end{figure}

In addition to the disk scale length and central surface brightness of the disk, we examine the effect of ignoring the break on the luminosity ratios such as B/T, D/T and Bar/T. In Figure~\ref{fig:comp_rscl5}, we show the differences of B/T, D/T, and Bar/T estimated from the fits that ignored and included the disk break. If the break is not accounted for, B/T and Bar/T ratios become smaller and D/T becomes larger compared to the fits when the break is included. The median with standard deviation of B/T is 0.91$\pm$0.13, Bar/T is 0.76$\pm$0.19 and D/T is 1.04$\pm$0.07 when the break is ignored. 

The reason why B/T and Bar/T increase when we introduce a break in the disk model is that the inner part of Type II disk is shallower than the outer one. Therefore ignoring a break and modeling a galaxy with a single exponential profile will result in flux being transferred from both bulge and the bar to the disk. As a result the model overestimates the disk in the inner part of the galaxy where the bulge and bar lie and the bulge and bar components are underestimated. For example, if we ignore the break and try to model galaxies, the Bar/T can be nearly 0 for some galaxies (bottom panel of Figure~\ref{fig:comp_rscl5}), even though there are clear bars in the 3.6 {\micron} image. 
While the impact on the total disk flux is not significant, it will have a large impact on the bulge and especially on the bar, underestimating their true luminosity/mass. This demonstrates that, although it is a complicating feature, accounting for disk breaks is critical for the study of the bulge and bar. It should be noted that there is a considerable dispersion in the differences in disk scale lengths and luminosity ratios so one cannot simply correct a fit without including the break using some a posterior rule.

\subsection{Implications}
Many studies use the disk scale length as a key measurement in studies of cosmic size evolution (e.g., \citealt{lilly_98, simard_99,barden_05,fathi_12}). 
However these studies ignore the presence of the disk break even though such breaks have been found out to redshift $\sim$ 1 (e.g., \citealt{perez_04, trujillo_05, azzollini_08}).  
Although we have compared how much the inner and outer disk scale lengths differ from the disk scale length from the fit that ignored the disk break, there is an important thing to consider. 
If a disk break is ignored, then the derived disk scale length changes depending on how much part of the disk is buried under the lifted sky, and the depth of the observation.
If most part of the outer disk is buried under the sky then the derived disk scale length preferentially represents inner part of the disk that is longer for Type II and shorter for Type III. However for nearby galaxies observed in depth and the outer disk is covered sufficiently, the derived disk scale length becomes shorter for Type II and longer for Type III. Therefore disk scale lengths obtained from galaxies observed with different limiting surface brightness may represent different parts of disks. 
Thus special care should be taken to avoid comparing disk scale lengths that represent different parts of disks.
This effect can be non-negligible in the study of the size evolution of disk galaxies using disk scale lengths to determine whether they scale with $H^{-1}(z)$ or $H^{-2/3}(z)$, where $H(z)$ is the Hubble parameter at redshift z.

Central surface brightness of disks is found to show a bimodal distribution (\citealt{sorce_13}) and two peaks are separated by 2 magnitude. However, the distribution of the central surface brightness will be changed if we include a break in the model fit. Our result shows that central surface brightness of galaxies decreases by $\sim$0.45 mag in median if disk breaks are properly modeled. Therefore we expect the distribution will be changed, but, the bimodality may still be in place.

\section{SUMMARY}

We have performed two-dimensional decomposition of 3.6 {\micron} images from the \s4g  for 144 nearby barred spiral galaxies.
Our sample covers various Hubble types and stellar masses from $10^9$ {\Msun} to $10^{11}$ \Msun.
Galaxies are decomposed with up to 4 subcomponents -- bulge, disk, bar, and a point source.
Because a majority of disk galaxies show at least one break in their radial profiles in the disk, we fit the disks with models that have a break. 
A one-page summary of our main results for each galaxy is given in Appendix~\ref{ap:summary}. These decompositions provide a number of important structural parameters, such as bulge effective radius and S\'ersic index, disk scale length, bar radius and bar S\'ersic index. 
We summarize our results as follows:

\begin{enumerate}

\item 
We find that $r_{\rm br}/R_{\rm bar}$ behave differently for galaxies with prominent bulges and for those with less prominent bulges. For galaxies with bulge-to-total (B/T) $\geq$ 0.1, $r_{\rm br}/R_{\rm bar}$ remains constant as the bulge becomes prominent. As a function of B/T, $r_{\rm br}/R_{\rm bar}$ forms a tighter sequence than as a function of Hubble types or stellar mass. However, galaxies with small bulges (B/T $<$ 0.1) exhibit a wide range of $r_{\rm br}/R_{\rm bar}$, with a slightly increased median $r_{\rm br}/R_{\rm bar}$.

\item
Different trends of $r_{\rm br}/R_{\rm bar}$ with B/T suggest that the mechanisms responsible for the break may be different for galaxies with B/T $<$ 0.1 and B/T $\geq$ 0.1. Breaks in galaxies with B/T$\geq$ 0.1 may be related to the Outer Lindblad Resonance (OLR) of the bar. As a bar grows, the break is also pushed out keeping $r_{\rm br}/R_{\rm bar}$ constant.
For galaxies with less conspicuous bulges, while the bar-only resonance may generate a break, other elements may be more efficient in creating breaks. Those elements include spiral-bar coupled resonances and a star formation threshold, which will increase and add scatter in the $r_{\rm br}/R_{\rm bar}$ distribution.

\item In Type II (down-bending) disk galaxies the median inner disk scale length is 2.21 times larger than the median outer disk scale length. Thus, if the break is ignored, then the derived disk scale length ($h$) becomes smaller than that of the inner disk ($h_{\rm in}$), and becomes larger than that of the outer one ($h_{\rm out}$). The median ratios are $h/h_{\rm in}$~$\sim$ 0.63 and $h/h_{\rm out}$~$\sim$1.38. Hence, it is important to model disk breaks in Type II disk galaxies to derive proper disk scale lengths. 

\item Modeling a galaxy without a break will result in a flux transfer from both bulge and bar to the disk, and thus underestimate their true luminosity.
If disk breaks are neglected for Type II disks, B/T and bar-to-total (Bar/T) luminosity ratios decrease by $\sim$ 10 and 25\% respectively in median, as compared to the fit that includes the break. Disk-to-total (D/T) ratio, however, increases by $\sim$ 5\%. 
Hence, to characterize both bar and bulge accurately, it is essential to account for a disk break.

\end{enumerate}

We have obtained structural parameters from sophisticated 2D decompositions, and there are many properties that we can explore to understand the formation and the evolution of barred galaxies. In a forthcoming paper, we will investigate the radial light profiles of bars to examine the relevance of the bulge prominence in this context, and explore the outer shape of bars to answer whether they vary with B/T or along the Hubble sequence. We will also revisit the bar length and ellipticity and their relation with B/T. Finally, we will explore the bar driven secular evolution of the disk.\\
 
The authors are grateful to the entire \s4g team for their effort in this project. 
We thank the anonymous referee for a thorough reading of the manuscript and helpful comments.
T.K., K.S., J.-C.M.-M., and T.M. acknowledge support from the National Radio Astronomy Observatory, which is a facility of the National Science Foundation operated under cooperative agreement by Associated Universities, Inc. We also gratefully acknowledge support from NASA JPL/Spitzer grant RSA 1374189 provided for the {\s4g} project.
T.K. and M.G.L. were supported by the National Research Foundation of Korea (NRF) grant funded by the Korea Government (MEST) (No. 2012R1A4A1028713).
D.A.G. thanks funding under the Marie Curie Actions of the European Commission (FP7-COFUND).
E.A. and A.B. thank the Centre National d'Etudes Spatiales for financial support. They also acknowledge financial support from the People Programme (Marie Curie Actions) of the European Union's FP7/2007-2013/ to the DAGAL network under REA grant agreement No. PITN-GA-2011-289313.
This research is based on observations and archival data made with the {\em{Spitzer}} Space Telescope, and made use of  the NASA/IPAC Extragalactic Database (NED) which are operated by the Jet Propulsion Laboratory, California Institute of Technology under a contract with National Aeronautics and Space Administration (NASA). We acknowledge the usage of the HyperLeda database (http://leda.univ-lyon1.fr).

\acknowledgments
{\it Facilities:} \facility{ The {\em{Spitzer}} Space Telescope}

\bibliographystyle{apj}
\bibliography{tkim_bar}  

\begin{thebibliography}{126}
\expandafter\ifx\csname natexlab\endcsname\relax\def\natexlab#1{#1}\fi

\bibitem[{{Allen} {et~al.}(2006){Allen}, {Driver}, {Graham}, {Cameron},
  {Liske}, \& {de Propris}}]{allen_06}
{Allen}, P.~D., {Driver}, S.~P., {Graham}, A.~W., {Cameron}, E., {Liske}, J.,
  \& {de Propris}, R. 2006, \mnras, 371, 2

\bibitem[{{Andredakis} {et~al.}(1995){Andredakis}, {Peletier}, \&
  {Balcells}}]{andredakis_95}
{Andredakis}, Y.~C., {Peletier}, R.~F., \& {Balcells}, M. 1995, \mnras, 275,
  874

\bibitem[{{Athanassoula}(1992)}]{athanassoula_92}
{Athanassoula}, E. 1992, \mnras, 259, 328

\bibitem[{{Athanassoula}(2003)}]{athanassoula_03}
---. 2003, \mnras, 341, 1179

\bibitem[{{Athanassoula}(2005)}]{athanassoula_05}
---. 2005, \mnras, 358, 1477

\bibitem[{{Athanassoula}(2012{\natexlab{a}})}]{athanassoula_12_canary}
---. 2012{\natexlab{a}}, ArXiv e-prints

\bibitem[{{Athanassoula}(2012{\natexlab{b}})}]{athanassoula_12_proc}
{Athanassoula}, E. 2012{\natexlab{b}}, in European Physical Journal Web of
  Conferences, Vol.~19, European Physical Journal Web of Conferences, 6004

\bibitem[{{Athanassoula} \& {Beaton}(2006)}]{athanassoula_06}
{Athanassoula}, E., \& {Beaton}, R.~L. 2006, \mnras, 370, 1499

\bibitem[{{Athanassoula} {et~al.}(1982){Athanassoula}, {Bosma}, {Creze}, \&
  {Schwarz}}]{athanassoula_82}
{Athanassoula}, E., {Bosma}, A., {Creze}, M., \& {Schwarz}, M.~P. 1982, \aap,
  107, 101

\bibitem[{{Athanassoula} {et~al.}(2013){Athanassoula}, {Machado}, \&
  {Rodionov}}]{athanassoula_13a}
{Athanassoula}, E., {Machado}, R.~E.~G., \& {Rodionov}, S.~A. 2013, \mnras,
  429, 1949

\bibitem[{{Athanassoula} \& {Misiriotis}(2002)}]{athanassoula_02a}
{Athanassoula}, E., \& {Misiriotis}, A. 2002, \mnras, 330, 35

\bibitem[{{Athanassoula} {et~al.}(1990){Athanassoula}, {Morin}, {Wozniak},
  {Puy}, {Pierce}, {Lombard}, \& {Bosma}}]{athanassoula_90}
{Athanassoula}, E., {Morin}, S., {Wozniak}, H., {Puy}, D., {Pierce}, M.~J.,
  {Lombard}, J., \& {Bosma}, A. 1990, \mnras, 245, 130

\bibitem[{{Athanassoula} {et~al.}(2009){Athanassoula}, {Romero-G{\'o}mez},
  {Bosma}, \& {Masdemont}}]{athanassoula_09b}
{Athanassoula}, E., {Romero-G{\'o}mez}, M., {Bosma}, A., \& {Masdemont}, J.~J.
  2009, \mnras, 400, 1706

\bibitem[{{Azzollini} {et~al.}(2008){Azzollini}, {Trujillo}, \&
  {Beckman}}]{azzollini_08}
{Azzollini}, R., {Trujillo}, I., \& {Beckman}, J.~E. 2008, \apj, 684, 1026

\bibitem[{{Bakos} {et~al.}(2008){Bakos}, {Trujillo}, \& {Pohlen}}]{bakos_08}
{Bakos}, J., {Trujillo}, I., \& {Pohlen}, M. 2008, \apjl, 683, L103

\bibitem[{{Barden} {et~al.}(2005){Barden}, {Rix}, {Somerville}, {Bell},
  {H{\"a}u{\ss}ler}, {Peng}, {Borch}, {Beckwith}, {Caldwell}, {Heymans},
  {Jahnke}, {Jogee}, {McIntosh}, {Meisenheimer}, {S{\'a}nchez}, {Wisotzki}, \&
  {Wolf}}]{barden_05}
{Barden}, M., {et~al.} 2005, \apj, 635, 959

\bibitem[{{Berentzen} {et~al.}(2007){Berentzen}, {Shlosman},
  {Martinez-Valpuesta}, \& {Heller}}]{berentzen_07}
{Berentzen}, I., {Shlosman}, I., {Martinez-Valpuesta}, I., \& {Heller}, C.~H.
  2007, \apj, 666, 189

\bibitem[{{Busch} {et~al.}(2013){Busch}, {Zuther}, {Valencia-S.}, {Moser},
  {Fischer}, {Eckart}, {Scharw{\"a}chter}, {Gadotti}, \&
  {Wisotzki}}]{busch_13xx}
{Busch}, G., {et~al.} 2013, ArXiv e-prints

\bibitem[{{Buta}(1995)}]{buta_95}
{Buta}, R. 1995, \apjs, 96, 39

\bibitem[{{Buta} \& {Combes}(1996)}]{buta_96}
{Buta}, R., \& {Combes}, F. 1996, \fcp, 17, 95

\bibitem[{{Buta} \& {Crocker}(1991)}]{buta_91}
{Buta}, R., \& {Crocker}, D.~A. 1991, \aj, 102, 1715

\bibitem[{{Buta} {et~al.}(2006){Buta}, {Laurikainen}, {Salo}, {Block}, \&
  {Knapen}}]{buta_06}
{Buta}, R., {Laurikainen}, E., {Salo}, H., {Block}, D.~L., \& {Knapen}, J.~H.
  2006, \aj, 132, 1859

\bibitem[{{Buta} {et~al.}(2010){Buta}, {Sheth}, {Regan}, {Hinz}, {Gil de Paz},
  {Men{\'e}ndez-Delmestre}, {Munoz-Mateos}, {Seibert}, {Laurikainen}, {Salo},
  {Gadotti}, {Athanassoula}, {Bosma}, {Knapen}, {Ho}, {Madore}, {Elmegreen},
  {Masters}, {Comer{\'o}n}, {Aravena}, \& {Kim}}]{buta_10}
{Buta}, R.~J., {et~al.} 2010, \apjs, 190, 147

\bibitem[{{Cameron} {et~al.}(2010){Cameron}, {Carollo}, {Oesch}, {Aller},
  {Bschorr}, {Cerulo}, {Aussel}, {Capak}, {Le Floc'h}, {Ilbert}, {Kneib},
  {Koekemoer}, {Leauthaud}, {Lilly}, {Massey}, {McCracken}, {Rhodes},
  {Salvato}, {Sanders}, {Scoville}, {Sheth}, {Taniguchi}, \&
  {Thompson}}]{cameron_10}
{Cameron}, E., {et~al.} 2010, \mnras, 409, 346

\bibitem[{{Caon} {et~al.}(1993){Caon}, {Capaccioli}, \& {D'Onofrio}}]{caon_93}
{Caon}, N., {Capaccioli}, M., \& {D'Onofrio}, M. 1993, \mnras, 265, 1013

\bibitem[{{Cheung} {et~al.}(2013){Cheung}, {Athanassoula}, {Masters}, {Nichol},
  {Bosma}, {Bell}, {Faber}, {Koo}, {Lintott}, {Melvin}, {Schawinski}, {Skibba},
  \& {Willett}}]{cheung_13xx}
{Cheung}, E., {et~al.} 2013, ArXiv e-prints

\bibitem[{{Combes} \& {Elmegreen}(1993)}]{combes_93}
{Combes}, F., \& {Elmegreen}, B.~G. 1993, \aap, 271, 391

\bibitem[{{Comer{\'o}n} {et~al.}(2010){Comer{\'o}n}, {Knapen}, {Beckman},
  {Laurikainen}, {Salo}, {Mart{\'{\i}}nez-Valpuesta}, \& {Buta}}]{comeron_10}
{Comer{\'o}n}, S., {Knapen}, J.~H., {Beckman}, J.~E., {Laurikainen}, E.,
  {Salo}, H., {Mart{\'{\i}}nez-Valpuesta}, I., \& {Buta}, R.~J. 2010, \mnras,
  402, 2462

\bibitem[{{Comer{\'o}n} {et~al.}(2011){Comer{\'o}n}, {Knapen}, {Sheth},
  {Regan}, {Hinz}, {Gil de Paz}, {Men{\'e}ndez-Delmestre}, {Mu{\~n}oz-Mateos},
  {Seibert}, {Kim}, {Athanassoula}, {Bosma}, {Buta}, {Elmegreen}, {Ho},
  {Holwerda}, {Laurikainen}, {Salo}, \& {Schinnerer}}]{comeron_11_4244}
{Comer{\'o}n}, S., {et~al.} 2011, \apj, 729, 18

\bibitem[{{Comer{\'o}n} {et~al.}(2012){Comer{\'o}n}, {Elmegreen}, {Salo},
  {Laurikainen}, {Athanassoula}, {Bosma}, {Knapen}, {Gadotti}, {Sheth}, {Hinz},
  {Regan}, {Gil de Paz}, {Mu{\~n}oz-Mateos}, {Men{\'e}ndez-Delmestre},
  {Seibert}, {Kim}, {Mizusawa}, {Laine}, {Ho}, \& {Holwerda}}]{comeron_12}
---. 2012, \apj, 759, 98

\bibitem[{{Comer{\'o}n} {et~al.}(2013){Comer{\'o}n}, {Salo}, {Laurikainen},
  {Knapen}, {Buta}, {Herrera-Endoqui}, {Laine}, {Holwerda}, {Sheth}, {Regan},
  {Hinz}, {Mu{\~n}oz-Mateoz}, {Gil de Paz}, {Men{\'e}ndez-Delmestre},
  {Seibert}, {Mizusawa}, {Kim}, {Erroz-Ferrer}, {Gadotti}, {Athanassoula},
  {Bosma}, \& {Ho}}]{comeron_13xx_ar}
---. 2013, ArXiv e-prints

\bibitem[{{de Jong}(1995)}]{dejong_95_t}
{de Jong}, R. 1995, PhD thesis, PhD thesis.~Univ.~Groningen , (1995)

\bibitem[{{de Jong} {et~al.}(2007){de Jong}, {Seth}, {Radburn-Smith}, {Bell},
  {Brown}, {Bullock}, {Courteau}, {Dalcanton}, {Ferguson}, {Goudfrooij},
  {Holfeltz}, {Holwerda}, {Purcell}, {Sick}, \& {Zucker}}]{dejong_07}
{de Jong}, R.~S., {et~al.} 2007, \apjl, 667, L49

\bibitem[{{de Souza} {et~al.}(2004){de Souza}, {Gadotti}, \& {dos
  Anjos}}]{desouza_04}
{de Souza}, R.~E., {Gadotti}, D.~A., \& {dos Anjos}, S. 2004, \apjs, 153, 411

\bibitem[{{de Vaucouleurs} {et~al.}(1991){de Vaucouleurs}, {de Vaucouleurs},
  {Corwin}, {Buta}, {Paturel}, \& {Fouqu{\'e}}}]{devaucouleurs_91}
{de Vaucouleurs}, G., {de Vaucouleurs}, A., {Corwin}, Jr., H.~G., {Buta},
  R.~J., {Paturel}, G., \& {Fouqu{\'e}}, P. 1991, {Third Reference Catalogue of
  Bright Galaxies. Volume I: Explanations and references. Volume II: Data for
  galaxies between 0$^{h}$ and 12$^{h}$. Volume III: Data for galaxies between
  12$^{h}$ and 24$^{h}$.}

\bibitem[{{Debattista} {et~al.}(2006){Debattista}, {Mayer}, {Carollo}, {Moore},
  {Wadsley}, \& {Quinn}}]{debattista_06}
{Debattista}, V.~P., {Mayer}, L., {Carollo}, C.~M., {Moore}, B., {Wadsley}, J.,
  \& {Quinn}, T. 2006, \apj, 645, 209

\bibitem[{{Debattista} \& {Sellwood}(2000)}]{debattista_00}
{Debattista}, V.~P., \& {Sellwood}, J.~A. 2000, \apj, 543, 704

\bibitem[{{D'Onofrio}(2001)}]{donofrio_01b}
{D'Onofrio}, M. 2001, \mnras, 326, 1517

\bibitem[{{Durbala} {et~al.}(2008){Durbala}, {Sulentic}, {Buta}, \&
  {Verdes-Montenegro}}]{durbala_08}
{Durbala}, A., {Sulentic}, J.~W., {Buta}, R., \& {Verdes-Montenegro}, L. 2008,
  \mnras, 390, 881

\bibitem[{{Elmegreen} \& {Elmegreen}(1985)}]{elmegreen_85}
{Elmegreen}, B.~G., \& {Elmegreen}, D.~M. 1985, \apj, 288, 438

\bibitem[{{Elmegreen} {et~al.}(2007){Elmegreen}, {Elmegreen}, {Knapen}, {Buta},
  {Block}, \& {Puerari}}]{elmegreen_07}
{Elmegreen}, B.~G., {Elmegreen}, D.~M., {Knapen}, J.~H., {Buta}, R.~J.,
  {Block}, D.~L., \& {Puerari}, I. 2007, \apjl, 670, L97

\bibitem[{{Elmegreen} \& {Hunter}(2006)}]{elmegreen_06}
{Elmegreen}, B.~G., \& {Hunter}, D.~A. 2006, \apj, 636, 712

\bibitem[{{Elmegreen} \& {Parravano}(1994)}]{elmegreen_94}
{Elmegreen}, B.~G., \& {Parravano}, A. 1994, \apjl, 435, L121

\bibitem[{{Erroz-Ferrer} {et~al.}(2012){Erroz-Ferrer}, {Knapen}, {Font},
  {Beckman}, {Falc{\'o}n-Barroso}, {S{\'a}nchez-Gallego}, {Athanassoula},
  {Bosma}, {Gadotti}, {Mu{\~n}oz-Mateos}, {Sheth}, {Buta}, {Comer{\'o}n}, {Gil
  de Paz}, {Hinz}, {Ho}, {Kim}, {Laine}, {Laurikainen}, {Madore},
  {Men{\'e}ndez-Delmestre}, {Mizusawa}, {Regan}, {Salo}, \&
  {Seibert}}]{erroz_ferrer_12}
{Erroz-Ferrer}, S., {et~al.} 2012, \mnras, 427, 2938

\bibitem[{{Erwin} {et~al.}(2005){Erwin}, {Beckman}, \&
  {Pohlen}}]{erwin_05_anti}
{Erwin}, P., {Beckman}, J.~E., \& {Pohlen}, M. 2005, \apjl, 626, L81

\bibitem[{{Erwin} \& {Debattista}(2013)}]{erwin_13}
{Erwin}, P., \& {Debattista}, V.~P. 2013, \mnras, 431, 3060

\bibitem[{{Erwin} \& {Gadotti}(2012)}]{erwin_12b}
{Erwin}, P., \& {Gadotti}, D.~A. 2012, Advances in Astronomy, 2012

\bibitem[{{Erwin} {et~al.}(2008){Erwin}, {Pohlen}, \& {Beckman}}]{erwin_08}
{Erwin}, P., {Pohlen}, M., \& {Beckman}, J.~E. 2008, \aj, 135, 20

\bibitem[{{Erwin} \& {Sparke}(2002)}]{erwin_02}
{Erwin}, P., \& {Sparke}, L.~S. 2002, \aj, 124, 65

\bibitem[{{Eskew} {et~al.}(2012){Eskew}, {Zaritsky}, \& {Meidt}}]{eskew_12}
{Eskew}, M., {Zaritsky}, D., \& {Meidt}, S. 2012, \aj, 143, 139

\bibitem[{{Eskridge} {et~al.}(2000){Eskridge}, {Frogel}, {Pogge}, {Quillen},
  {Davies}, {DePoy}, {Houdashelt}, {Kuchinski}, {Ram{\'{\i}}rez}, {Sellgren},
  {Terndrup}, \& {Tiede}}]{eskridge_00}
{Eskridge}, P.~B., {et~al.} 2000, \aj, 119, 536

\bibitem[{{Fathi} {et~al.}(2012){Fathi}, {Gatchell}, {Hatziminaoglou}, \&
  {Epinat}}]{fathi_12}
{Fathi}, K., {Gatchell}, M., {Hatziminaoglou}, E., \& {Epinat}, B. 2012,
  \mnras, 423, L112

\bibitem[{{Fazio} {et~al.}(2004){Fazio}, {Hora}, {Allen}, {Ashby}, {Barmby},
  {Deutsch}, {Huang}, {Kleiner}, {Marengo}, {Megeath}, {Melnick}, {Pahre},
  {Patten}, {Polizotti}, {Smith}, {Taylor}, {Wang}, {Willner}, {Hoffmann},
  {Pipher}, {Forrest}, {McMurty}, {McCreight}, {McKelvey}, {McMurray}, {Koch},
  {Moseley}, {Arendt}, {Mentzell}, {Marx}, {Losch}, {Mayman}, {Eichhorn},
  {Krebs}, {Jhabvala}, {Gezari}, {Fixsen}, {Flores}, {Shakoorzadeh}, {Jungo},
  {Hakun}, {Workman}, {Karpati}, {Kichak}, {Whitley}, {Mann}, {Tollestrup},
  {Eisenhardt}, {Stern}, {Gorjian}, {Bhattacharya}, {Carey}, {Nelson},
  {Glaccum}, {Lacy}, {Lowrance}, {Laine}, {Reach}, {Stauffer}, {Surace},
  {Wilson}, {Wright}, {Hoffman}, {Domingo}, \& {Cohen}}]{fazio_04}
{Fazio}, G.~G., {et~al.} 2004, \apjs, 154, 10

\bibitem[{{Foyle} {et~al.}(2008){Foyle}, {Courteau}, \& {Thacker}}]{foyle_08}
{Foyle}, K., {Courteau}, S., \& {Thacker}, R.~J. 2008, \mnras, 386, 1821

\bibitem[{{Freeman}(1970)}]{freeman_70}
{Freeman}, K.~C. 1970, \apj, 160, 811

\bibitem[{{Gadotti}(2008)}]{gadotti_08}
{Gadotti}, D.~A. 2008, \mnras, 384, 420

\bibitem[{{Gadotti}(2009)}]{gadotti_09}
---. 2009, \mnras, 393, 1531

\bibitem[{{Gadotti}(2011)}]{gadotti_11}
---. 2011, \mnras, 415, 3308

\bibitem[{{Gadotti} {et~al.}(2007){Gadotti}, {Athanassoula}, {Carrasco},
  {Bosma}, {de Souza}, \& {Recillas}}]{gadotti_07}
{Gadotti}, D.~A., {Athanassoula}, E., {Carrasco}, L., {Bosma}, A., {de Souza},
  R.~E., \& {Recillas}, E. 2007, \mnras, 381, 943

\bibitem[{{Graham}(2001)}]{graham_01}
{Graham}, A.~W. 2001, \aj, 121, 820

\bibitem[{{Guti{\'e}rrez} {et~al.}(2011){Guti{\'e}rrez}, {Erwin}, {Aladro}, \&
  {Beckman}}]{gutierrez_11}
{Guti{\'e}rrez}, L., {Erwin}, P., {Aladro}, R., \& {Beckman}, J.~E. 2011, \aj,
  142, 145

\bibitem[{{H{\"a}ussler} {et~al.}(2007){H{\"a}ussler}, {McIntosh}, {Barden},
  {Bell}, {Rix}, {Borch}, {Beckwith}, {Caldwell}, {Heymans}, {Jahnke}, {Jogee},
  {Koposov}, {Meisenheimer}, {S{\'a}nchez}, {Somerville}, {Wisotzki}, \&
  {Wolf}}]{haussler_07}
{H{\"a}ussler}, B., {et~al.} 2007, \apjs, 172, 615

\bibitem[{{Hohl}(1971)}]{hohl_71}
{Hohl}, F. 1971, \apj, 168, 343

\bibitem[{{Huang} {et~al.}(2013){Huang}, {Ho}, {Peng}, {Li}, \&
  {Barth}}]{huang_13}
{Huang}, S., {Ho}, L.~C., {Peng}, C.~Y., {Li}, Z.-Y., \& {Barth}, A.~J. 2013,
  \apj, 766, 47

\bibitem[{{Huertas-Company} {et~al.}(2007){Huertas-Company}, {Rouan},
  {Soucail}, {Le F{\`e}vre}, {Tasca}, \& {Contini}}]{huertas_company_07}
{Huertas-Company}, M., {Rouan}, D., {Soucail}, G., {Le F{\`e}vre}, O., {Tasca},
  L., \& {Contini}, T. 2007, \aap, 468, 937

\bibitem[{{Hunter} \& {Elmegreen}(2006)}]{hunter_06}
{Hunter}, D.~A., \& {Elmegreen}, B.~G. 2006, \apjs, 162, 49

\bibitem[{{Kalnajs}(1972)}]{kalnajs_72}
{Kalnajs}, A.~J. 1972, \apj, 175, 63

\bibitem[{{Kazantzidis} {et~al.}(2009){Kazantzidis}, {Zentner}, {Kravtsov},
  {Bullock}, \& {Debattista}}]{kazantzidis_09}
{Kazantzidis}, S., {Zentner}, A.~R., {Kravtsov}, A.~V., {Bullock}, J.~S., \&
  {Debattista}, V.~P. 2009, \apj, 700, 1896

\bibitem[{{Kennicutt}(1989)}]{kennicutt_89}
{Kennicutt}, Jr., R.~C. 1989, \apj, 344, 685

\bibitem[{{Khosroshahi} {et~al.}(2000){Khosroshahi}, {Wadadekar}, \&
  {Kembhavi}}]{khoroshahi_00}
{Khosroshahi}, H.~G., {Wadadekar}, Y., \& {Kembhavi}, A. 2000, \apj, 533, 162

\bibitem[{{Kim} {et~al.}(2012){Kim}, {Sheth}, {Hinz}, {Lee}, {Zaritsky},
  {Gadotti}, {Knapen}, {Schinnerer}, {Ho}, {Laurikainen}, {Salo},
  {Athanassoula}, {Bosma}, {de Swardt}, {Mu{\~n}oz-Mateos}, {Madore},
  {Comer{\'o}n}, {Regan}, {Men{\'e}ndez-Delmestre}, {Gil de Paz}, {Seibert},
  {Laine}, {Erroz-Ferrer}, \& {Mizusawa}}]{kim_12}
{Kim}, T., {et~al.} 2012, \apj, 753, 43

\bibitem[{{Kormendy}(1979)}]{kormendy_79}
{Kormendy}, J. 1979, \apj, 227, 714

\bibitem[{{Laurikainen} \& {Salo}(2001)}]{laurikainen_01}
{Laurikainen}, E., \& {Salo}, H. 2001, \mnras, 324, 685

\bibitem[{{Laurikainen} {et~al.}(2013){Laurikainen}, {Salo}, {Athanassoula},
  {Bosma}, {Buta}, \& {Janz}}]{laurikainen_13}
{Laurikainen}, E., {Salo}, H., {Athanassoula}, E., {Bosma}, A., {Buta}, R., \&
  {Janz}, J. 2013, \mnras, 430, 3489

\bibitem[{{Laurikainen} {et~al.}(2004){Laurikainen}, {Salo}, \&
  {Buta}}]{laurikainen_04_bar}
{Laurikainen}, E., {Salo}, H., \& {Buta}, R. 2004, \apj, 607, 103

\bibitem[{{Laurikainen} {et~al.}(2005){Laurikainen}, {Salo}, \&
  {Buta}}]{laurikainen_05}
---. 2005, \mnras, 362, 1319

\bibitem[{{Laurikainen} {et~al.}(2006){Laurikainen}, {Salo}, {Buta}, {Knapen},
  {Speltincx}, \& {Block}}]{laurikainen_06}
{Laurikainen}, E., {Salo}, H., {Buta}, R., {Knapen}, J., {Speltincx}, T., \&
  {Block}, D. 2006, \aj, 132, 2634

\bibitem[{{Laurikainen} {et~al.}(2007){Laurikainen}, {Salo}, {Buta}, \&
  {Knapen}}]{laurikainen_07}
{Laurikainen}, E., {Salo}, H., {Buta}, R., \& {Knapen}, J.~H. 2007, \mnras,
  381, 401

\bibitem[{{Laurikainen} {et~al.}(2009){Laurikainen}, {Salo}, {Buta}, \&
  {Knapen}}]{laurikainen_09}
---. 2009, \apjl, 692, L34

\bibitem[{{Laurikainen} {et~al.}(2010){Laurikainen}, {Salo}, {Buta}, {Knapen},
  \& {Comer{\'o}n}}]{laurikainen_10}
{Laurikainen}, E., {Salo}, H., {Buta}, R., {Knapen}, J.~H., \& {Comer{\'o}n},
  S. 2010, \mnras, 405, 1089

\bibitem[{{Lilly} {et~al.}(1998){Lilly}, {Schade}, {Ellis}, {Le Fevre},
  {Brinchmann}, {Tresse}, {Abraham}, {Hammer}, {Crampton}, {Colless},
  {Glazebrook}, {Mallen-Ornelas}, \& {Broadhurst}}]{lilly_98}
{Lilly}, S., {et~al.} 1998, \apj, 500, 75

\bibitem[{{Little} \& {Carlberg}(1991)}]{little_91a}
{Little}, B., \& {Carlberg}, R.~G. 1991, \mnras, 250, 161

\bibitem[{{Maltby} {et~al.}(2012{\natexlab{a}}){Maltby}, {Hoyos}, {Gray},
  {Arag{\'o}n-Salamanca}, \& {Wolf}}]{maltby_12b}
{Maltby}, D.~T., {Hoyos}, C., {Gray}, M.~E., {Arag{\'o}n-Salamanca}, A., \&
  {Wolf}, C. 2012{\natexlab{a}}, \mnras, 420, 2475

\bibitem[{{Maltby} {et~al.}(2012{\natexlab{b}}){Maltby}, {Gray},
  {Arag{\'o}n-Salamanca}, {Wolf}, {Bell}, {Jogee}, {H{\"a}u{\ss}ler},
  {Barazza}, {B{\"o}hm}, \& {Jahnke}}]{maltby_12a}
{Maltby}, D.~T., {et~al.} 2012{\natexlab{b}}, \mnras, 419, 669

\bibitem[{{Marleau} \& {Simard}(1998)}]{marleau_98}
{Marleau}, F.~R., \& {Simard}, L. 1998, \apj, 507, 585

\bibitem[{{Mart{\'{\i}}n-Navarro} {et~al.}(2012){Mart{\'{\i}}n-Navarro},
  {Bakos}, {Trujillo}, {Knapen}, {Athanassoula}, {Bosma}, {Comer{\'o}n},
  {Elmegreen}, {Erroz-Ferrer}, {Gadotti}, {Gil de Paz}, {Hinz}, {Ho},
  {Holwerda}, {Kim}, {Laine}, {Laurikainen}, {Men{\'e}ndez-Delmestre},
  {Mizusawa}, {Mu{\~n}oz-Mateos}, {Regan}, {Salo}, {Seibert}, \&
  {Sheth}}]{martin_navarro_12}
{Mart{\'{\i}}n-Navarro}, I., {et~al.} 2012, \mnras, 427, 1102

\bibitem[{{Martinez-Valpuesta} {et~al.}(2006){Martinez-Valpuesta}, {Shlosman},
  \& {Heller}}]{martinez_valpuesta_06}
{Martinez-Valpuesta}, I., {Shlosman}, I., \& {Heller}, C. 2006, \apj, 637, 214

\bibitem[{{Meidt} {et~al.}(2012{\natexlab{a}}){Meidt}, {Schinnerer}, {Knapen},
  {Bosma}, {Athanassoula}, {Sheth}, {Buta}, {Zaritsky}, {Laurikainen},
  {Elmegreen}, {Elmegreen}, {Gadotti}, {Salo}, {Regan}, {Ho}, {Madore}, {Hinz},
  {Skibba}, {Gil de Paz}, {Mu{\~n}oz-Mateos}, {Men{\'e}ndez-Delmestre},
  {Seibert}, {Kim}, {Mizusawa}, {Laine}, \& {Comer{\'o}n}}]{meidt_12a}
{Meidt}, S.~E., {et~al.} 2012{\natexlab{a}}, \apj, 744, 17

\bibitem[{{Meidt} {et~al.}(2012{\natexlab{b}}){Meidt}, {Schinnerer},
  {Mu{\~n}oz-Mateos}, {Holwerda}, {Ho}, {Madore}, {Knapen}, {Bosma},
  {Athanassoula}, {Hinz}, {Sheth}, {Regan}, {Gil de Paz},
  {Men{\'e}ndez-Delmestre}, {Seibert}, {Kim}, {Mizusawa}, {Gadotti},
  {Laurikainen}, {Salo}, {Laine}, \& {Comer{\'o}n}}]{meidt_12b}
---. 2012{\natexlab{b}}, \apjl, 748, L30

\bibitem[{{Men{\'e}ndez-Delmestre} {et~al.}(2007){Men{\'e}ndez-Delmestre},
  {Sheth}, {Schinnerer}, {Jarrett}, \& {Scoville}}]{menendez_delmestre_07}
{Men{\'e}ndez-Delmestre}, K., {Sheth}, K., {Schinnerer}, E., {Jarrett}, T.~H.,
  \& {Scoville}, N.~Z. 2007, \apj, 657, 790

\bibitem[{{Minchev} {et~al.}(2012){Minchev}, {Famaey}, {Quillen}, {Di Matteo},
  {Combes}, {Vlaji{\'c}}, {Erwin}, \& {Bland-Hawthorn}}]{minchev_12a}
{Minchev}, I., {Famaey}, B., {Quillen}, A.~C., {Di Matteo}, P., {Combes}, F.,
  {Vlaji{\'c}}, M., {Erwin}, P., \& {Bland-Hawthorn}, J. 2012, \aap, 548, A126

\bibitem[{{M{\"o}llenhoff} \& {Heidt}(2001)}]{mollenhoff_01}
{M{\"o}llenhoff}, C., \& {Heidt}, J. 2001, \aap, 368, 16

\bibitem[{{Mu{\~n}oz-Mateos} {et~al.}(2013){Mu{\~n}oz-Mateos}, {Sheth}, {Gil de
  Paz}, {Meidt}, {Athanassoula}, {Bosma}, {Comer{\'o}n}, {Elmegreen},
  {Elmegreen}, {Erroz-Ferrer}, {Gadotti}, {Hinz}, {Ho}, {Holwerda}, {Jarrett},
  {Kim}, {Knapen}, {Laine}, {Laurikainen}, {Madore}, {Menendez-Delmestre},
  {Mizusawa}, {Regan}, {Salo}, {Schinnerer}, {Seibert}, {Skibba}, \&
  {Zaritsky}}]{munoz_mateos_13}
{Mu{\~n}oz-Mateos}, J.~C., {et~al.} 2013, \apj, 771, 59

\bibitem[{{Ostriker} \& {Peebles}(1973)}]{ostriker_73}
{Ostriker}, J.~P., \& {Peebles}, P.~J.~E. 1973, \apj, 186, 467

\bibitem[{{Paturel} {et~al.}(2003){Paturel}, {Petit}, {Prugniel}, {Theureau},
  {Rousseau}, {Brouty}, {Dubois}, \& {Cambr{\'e}sy}}]{paturel_03}
{Paturel}, G., {Petit}, C., {Prugniel}, P., {Theureau}, G., {Rousseau}, J.,
  {Brouty}, M., {Dubois}, P., \& {Cambr{\'e}sy}, L. 2003, \aap, 412, 45

\bibitem[{{Peng} {et~al.}(2002){Peng}, {Ho}, {Impey}, \& {Rix}}]{peng_02}
{Peng}, C.~Y., {Ho}, L.~C., {Impey}, C.~D., \& {Rix}, H.-W. 2002, \aj, 124, 266

\bibitem[{{Peng} {et~al.}(2010){Peng}, {Ho}, {Impey}, \& {Rix}}]{peng_10}
---. 2010, \aj, 139, 2097

\bibitem[{{P{\'e}rez}(2004)}]{perez_04}
{P{\'e}rez}, I. 2004, \aap, 427, L17

\bibitem[{{Pignatelli} {et~al.}(2006){Pignatelli}, {Fasano}, \&
  {Cassata}}]{pignatelli_06}
{Pignatelli}, E., {Fasano}, G., \& {Cassata}, P. 2006, \aap, 446, 373

\bibitem[{{Pohlen} {et~al.}(2002){Pohlen}, {Dettmar}, {L{\"u}tticke}, \&
  {Aronica}}]{pohlen_02}
{Pohlen}, M., {Dettmar}, R.-J., {L{\"u}tticke}, R., \& {Aronica}, G. 2002,
  \aap, 392, 807

\bibitem[{{Pohlen} \& {Trujillo}(2006)}]{pohlen_06}
{Pohlen}, M., \& {Trujillo}, I. 2006, \aap, 454, 759

\bibitem[{{Quillen} {et~al.}(2011){Quillen}, {Dougherty}, {Bagley}, {Minchev},
  \& {Comparetta}}]{quillen_11}
{Quillen}, A.~C., {Dougherty}, J., {Bagley}, M.~B., {Minchev}, I., \&
  {Comparetta}, J. 2011, \mnras, 417, 762

\bibitem[{{Radburn-Smith} {et~al.}(2012){Radburn-Smith}, {Ro{\v s}kar},
  {Debattista}, {Dalcanton}, {Streich}, {de Jong}, {Vlaji{\'c}}, {Holwerda},
  {Purcell}, {Dolphin}, \& {Zucker}}]{radburn_smith_12}
{Radburn-Smith}, D.~J., {et~al.} 2012, \apj, 753, 138

\bibitem[{{Rautiainen} \& {Salo}(1999)}]{rautiainen_99}
{Rautiainen}, P., \& {Salo}, H. 1999, \aap, 348, 737

\bibitem[{{Rautiainen} {et~al.}(2005){Rautiainen}, {Salo}, \&
  {Laurikainen}}]{rautiainen_05}
{Rautiainen}, P., {Salo}, H., \& {Laurikainen}, E. 2005, \apjl, 631, L129

\bibitem[{{Rautiainen} {et~al.}(2008){Rautiainen}, {Salo}, \&
  {Laurikainen}}]{rautiainen_08}
---. 2008, \mnras, 388, 1803

\bibitem[{{Ro{\v s}kar} {et~al.}(2008){Ro{\v s}kar}, {Debattista}, {Stinson},
  {Quinn}, {Kaufmann}, \& {Wadsley}}]{roskar_08a}
{Ro{\v s}kar}, R., {Debattista}, V.~P., {Stinson}, G.~S., {Quinn}, T.~R.,
  {Kaufmann}, T., \& {Wadsley}, J. 2008, \apjl, 675, L65

\bibitem[{{S{\'a}nchez-Bl{\'a}zquez} {et~al.}(2009){S{\'a}nchez-Bl{\'a}zquez},
  {Courty}, {Gibson}, \& {Brook}}]{sanchez_blazquez_09}
{S{\'a}nchez-Bl{\'a}zquez}, P., {Courty}, S., {Gibson}, B.~K., \& {Brook},
  C.~B. 2009, \mnras, 398, 591

\bibitem[{{Schaye}(2004)}]{schaye_04}
{Schaye}, J. 2004, \apj, 609, 667

\bibitem[{{Schwarz}(1981)}]{schwarz_81}
{Schwarz}, M.~P. 1981, \apj, 247, 77

\bibitem[{{S{\'e}rsic}(1963)}]{sersic_63}
{S{\'e}rsic}, J.~L. 1963, Boletin de la Asociacion Argentina de Astronomia La
  Plata Argentina, 6, 41

\bibitem[{{Sheth} {et~al.}(2012){Sheth}, {Melbourne}, {Elmegreen}, {Elmegreen},
  {Athanassoula}, {Abraham}, \& {Weiner}}]{sheth_12}
{Sheth}, K., {Melbourne}, J., {Elmegreen}, D.~M., {Elmegreen}, B.~G.,
  {Athanassoula}, E., {Abraham}, R.~G., \& {Weiner}, B.~J. 2012, \apj, 758, 136

\bibitem[{{Sheth} {et~al.}(2004){Sheth}, {Menendez-Delmestre}, {Scoville},
  {Jarrett}, {Strubbe}, {Regan}, {Schinnerer}, \& {Block}}]{sheth_04}
{Sheth}, K., {Menendez-Delmestre}, K., {Scoville}, N., {Jarrett}, T.,
  {Strubbe}, L., {Regan}, M.~W., {Schinnerer}, E., \& {Block}, D.~L. 2004, in
  Astrophysics and Space Science Library, Vol. 319, Penetrating Bars Through
  Masks of Cosmic Dust, ed. D.~L. {Block}, I.~{Puerari}, K.~C. {Freeman},
  R.~{Groess}, \& E.~K. {Block}, 405

\bibitem[{{Sheth} {et~al.}(2000){Sheth}, {Regan}, {Vogel}, \&
  {Teuben}}]{sheth_00}
{Sheth}, K., {Regan}, M.~W., {Vogel}, S.~N., \& {Teuben}, P.~J. 2000, \apj,
  532, 221

\bibitem[{{Sheth} {et~al.}(2002){Sheth}, {Vogel}, {Regan}, {Teuben}, {Harris},
  \& {Thornley}}]{sheth_02}
{Sheth}, K., {Vogel}, S.~N., {Regan}, M.~W., {Teuben}, P.~J., {Harris}, A.~I.,
  \& {Thornley}, M.~D. 2002, \aj, 124, 2581

\bibitem[{{Sheth} {et~al.}(2008){Sheth}, {Elmegreen}, {Elmegreen}, {Capak},
  {Abraham}, {Athanassoula}, {Ellis}, {Mobasher}, {Salvato}, {Schinnerer},
  {Scoville}, {Spalsbury}, {Strubbe}, {Carollo}, {Rich}, \& {West}}]{sheth_08}
{Sheth}, K., {et~al.} 2008, \apj, 675, 1141

\bibitem[{{Sheth} {et~al.}(2010){Sheth}, {Regan}, {Hinz}, {Gil de Paz},
  {Men{\'e}ndez-Delmestre}, {Mu{\~n}oz-Mateos}, {Seibert}, {Kim},
  {Laurikainen}, {Salo}, {Gadotti}, {Laine}, {Mizusawa}, {Armus},
  {Athanassoula}, {Bosma}, {Buta}, {Capak}, {Jarrett}, {Elmegreen},
  {Elmegreen}, {Knapen}, {Koda}, {Helou}, {Ho}, {Madore}, {Masters},
  {Mobasher}, {Ogle}, {Peng}, {Schinnerer}, {Surace}, {Zaritsky},
  {Comer{\'o}n}, {de Swardt}, {Meidt}, {Kasliwal}, \& {Aravena}}]{sheth_10}
---. 2010, \pasp, 122, 1397

\bibitem[{{Simard} {et~al.}(1999){Simard}, {Koo}, {Faber}, {Sarajedini},
  {Vogt}, {Phillips}, {Gebhardt}, {Illingworth}, \& {Wu}}]{simard_99}
{Simard}, L., {et~al.} 1999, \apj, 519, 563

\bibitem[{{Sorce} {et~al.}(2013){Sorce}, {Courtois}, {Sheth}, \&
  {Tully}}]{sorce_13}
{Sorce}, J.~G., {Courtois}, H.~M., {Sheth}, K., \& {Tully}, R.~B. 2013, \mnras,
  433, 751

\bibitem[{{Sygnet} {et~al.}(1988){Sygnet}, {Tagger}, {Athanassoula}, \&
  {Pellat}}]{sygnet_88}
{Sygnet}, J.~F., {Tagger}, M., {Athanassoula}, E., \& {Pellat}, R. 1988,
  \mnras, 232, 733

\bibitem[{{Tagger} {et~al.}(1987){Tagger}, {Sygnet}, {Athanassoula}, \&
  {Pellat}}]{tagger_87}
{Tagger}, M., {Sygnet}, J.~F., {Athanassoula}, E., \& {Pellat}, R. 1987, \apjl,
  318, L43

\bibitem[{{Trujillo} \& {Pohlen}(2005)}]{trujillo_05}
{Trujillo}, I., \& {Pohlen}, M. 2005, \apjl, 630, L17

\bibitem[{{van der Kruit}(1979)}]{vanderkruit_79}
{van der Kruit}, P.~C. 1979, \aaps, 38, 15

\bibitem[{{van der Kruit}(1987)}]{vanderkruit_87}
---. 1987, \aap, 173, 59

\bibitem[{{van der Kruit} \& {Searle}(1981)}]{vanderkruit_81a}
{van der Kruit}, P.~C., \& {Searle}, L. 1981, \aap, 95, 105

\bibitem[{{Younger} {et~al.}(2007){Younger}, {Cox}, {Seth}, \&
  {Hernquist}}]{younger_07}
{Younger}, J.~D., {Cox}, T.~J., {Seth}, A.~C., \& {Hernquist}, L. 2007, \apj,
  670, 269

\end{thebibliography}

\clearpage
\appendix
\section{A. Comparisons with other studies}
\subsection{A.1. Structural parameters from different methods}
\label{ap:diff_method}
\begin{figure*}]
\centering
\includegraphics[keepaspectratio=true,width=17.5cm,clip=true]{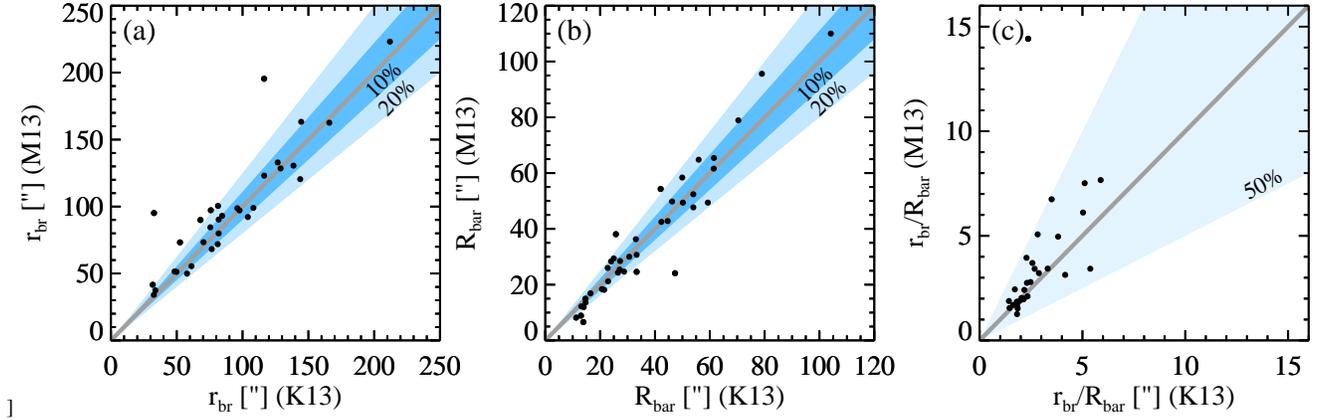}
\caption{Comparison of (a) break radii ($r_{\rm br}$), (b) deprojected bar radii ($R_{\rm bar}$), and (c) $r_{\rm br}$/$R_{\rm bar}$ estimated from this study (K13) and Mu{\~n}oz-Mateos et al. (2013, M13). Grey line indicates one-to-one line for comparison and shaded regions cover 10, 20, and 50\% deviation from one-to-one line, as indicated.
}
\label{fig:comp_jcmm}
\end{figure*}

In this study we estimate structural parameters from 2D model fit, while most other studies on disk break are from azimuthally averaged radial profile. 
However, estimated structural parameters may differ depending on adopted methods.
Therefore to evaluate differences of structural parameters from different methods, we compare $r_{\rm br}$ and $R_{\rm bar}$ estimated from this study and those from \citet{munoz_mateos_13} in Figure~\ref{fig:comp_jcmm}. There are 88 galaxies in common, but after {\bf (i)} removing galaxies that have a break at an inner ring or a lens, {\bf (ii)} removing galaxies classified as Type II.i by \citet{munoz_mateos_13} and thus break radii were not presented, and {\bf (iii)} choosing galaxies which both studies identified only one Type II disk break per galaxy, we are left with 32 galaxies and those are plotted in Figure~\ref{fig:comp_jcmm}. Measurements of break and bar radii from two studies agree within 20\% except for a few galaxies. Figure~\ref{fig:comp_jcmm} (c), however, shows that the ratio of the two ($r_{\rm br}/R_{\rm bar}$) differs by up to 50\%. So while the break and bar radii measurements are in agreement, the ratio of the two differs significantly in some cases. This shows that different methods of measuring structural properties of galaxies can give us different results. Thus it is important to keep in mind such differences when extrapolating results of the measurements and/or gathering results from different studies. 


\subsection{A.2. Connection between Breaks and Outer Rings}
\label{ap:outering}
Outer rings, whether they are closed authentic rings, or tightly-wound spiral arms are thought to be closely related to the OLR (\citealt{schwarz_81}). They are located roughly at the radius of around twice the bar radius (\citealt{kormendy_79, athanassoula_82}). \citet{buta_95} published the Catalogue of Southern Ring Galaxies and evaluated whether rings are related to orbital resonances with a bar or oval. He showed that outer rings are likely tracers of the location of the OLR, while inner rings are due to the ultra harmonic resonance (See also \citealt{buta_91}). 

Outer rings can cause a change in the slope of disk profiles. Indeed, many breaks occur at twice the bar radius (\citealt{erwin_08, munoz_mateos_13}), as we confirm here. Therefore some breaks are likely to be associated with outer rings. It is worth reviewing how rings populate along the Hubble sequence. \citet{buta_96} find that $\sim$ 10\% of disk galaxies exhibit outer rings, and $\sim$45\% of disk galaxies have inner rings based on the Third Reference Catalogue of Bright Galaxies (\citealt{devaucouleurs_91}). \citet{comeron_13xx_ar} conduct a statistical study on 724 ring galaxies (including pseudo ring) from \s4g and present the atlas, size, and frequency of resonance rings (ARRAKIS: Atlas of Resonance Rings as Known In the \s4g). They show that the outer and inner rings account for 16 $\pm$ 1 and 35 $\pm$ 1\%, respectively. 
They also find that outer rings are 1.7 times more common among barred galaxies than among unbarred galaxies, while inner rings are 1.3 times more common among barred galaxies than among unbarred galaxies (also see results of \citealt{laurikainen_13} from Near-Infrared S0 Survey).
Interestingly, they find that outer rings are mostly in Hubble stages $-1$ $\leq$ T $\leq$ 4 while inner rings are distributed broadly covering  $-1$ $\leq$ T $\leq$ $7$. As outer rings are usually thought to be limited by the OLR of the bar (\citealt{schwarz_81, athanassoula_09b}), this is consistent with our result that galaxies with B/T $\geq$ 0.1 (early type disks) have constant $r_{\rm br}/R_{\rm Bar}$ and exhibit less scatter of $r_{\rm br}/R_{\rm Bar}$ than galaxies with B/T $<$ 0.1 (late type disks) do.
We examine whether the location of the break is related to the outer ring, which has been pointed out by \citet{pohlen_06}, and \citet{erwin_08}. Using MID-IR classification (\citealt{buta_10}; R. Buta et al 2013, in prep) that is also presented in Table 1 and outer ring radii data from ARRAKIS (\citealt{comeron_13xx_ar}), out of 144 galaxies, we find 60 galaxies that are classified to have an outer ring (including pseudo outer ring and outer ring lens), and  whose outer ring radii are presented.

\begin{figure}
\centering
  \vskip 0.3cm
\includegraphics[keepaspectratio=true,width=8cm,clip=true]{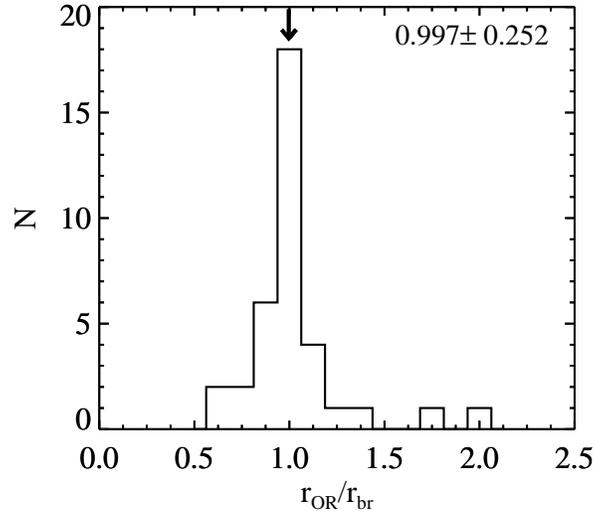}
\caption{Distribution of the outer ring radius ($r_{\rm OR}$) to break radius ($r_{\rm br}$). Galaxies that are classified to have outer ring, pseudo outer ring, or outer ring lens are included. $r_{\rm OR}$ is taken from \citet{comeron_13xx_ar}. The median and the standard deviation of the distribution are presented on the upper right corner, and the arrow indicates the median of the distribution.
}
\label{fig:ror_rbr}
\end{figure}

Among them, there are 24 galaxies in which we find the break at the edge of inner ring or lens. 
We examine the other 36 galaxies and plot the distribution of the outer ring radius ($r_{\rm OR}$) to break radius ($r_{\rm br}$) in Figure~\ref{fig:ror_rbr}.
We find that 83\% (30/36) of the outer ring galaxies in our sample show outer rings between 0.8$\times r_{\rm br}$ and 1.2$\times r_{\rm br}$. In particular, NGC 210 and NGC 5101 have two outer rings (including outer ring lens) in those region. 
If we confine to the outer rings that arise between 0.9$\times r_{\rm br}$ and 1.1$\times r_{\rm br}$,  we find that 64\% of ring galaxies (23/36) possess an outer ring at the break. Those galaxies have $r_{\rm br}/R_{\rm bar} \sim $1--3.5, and 0$\leq$B/T$<$0.5. 
Although we find breaks at the edge of inner rings or lenses for some galaxies, such galaxies may also have a second break at the outer ring.
Therefore we visually examine surface brightness profiles of such galaxies and find that additionally 12 out of 24 galaxies with an inner ring or a lens have another break at the outer (pseudo) ring.  

To summarize, out of all barred galaxies with an outer ring (60 galaxies), we find that more than half of them ($\sim$70\% for 0.8$\leq r_{\rm OR}/r{\rm_{br}} \leq$1.2, and $\sim$58\% for 0.9$\leq r_{\rm OR}/r{\rm_{br}} \leq$1.1 ) show a break at the outer ring. 
However, $\sim$ 8\% (5/60) of our sample barred galaxies with an outer ring do not show a break in their radial profile.
Also there are 9 galaxies\footnote{NGC 210, NGC 1302, NGC 1367, NGC 4579, NGC 4984, NGC 5101, NGC 7051, NGC 7731, and PGC053093.} that have two outer rings or an outer ring with an outer ring lens (\citealt{buta_10}, R. Buta et al. 2013, in prep). Half of them show a break at the outer-outer ring, however for the other half of them, we find the break at another position. 

J. Laine et al. (2013, MNRAS submitted) also find that break radii of about half of the Type II disk galaxies are associated with ring features (including outer rings, outer pseudo rings, and outer ringlenses). In particular, they find that breaks in earlier Hubble type disk galaxies (T$<$3) are closely related to ring features.

Outer rings are thought to be located at resonances, therefore our results imply that breaks also arise at the resonance. There is a possibility that outer rings would cause the break photometrically. However, the outer ring may cause the break physically. As inner disk matter beyond the $R_{\rm CR}$ are pushed outward, matters are piled up at the outer ring, which would make inner disk flatter and the outer edge of the outer ring falls steep. Therefore as the outer ring forms, the inner disk becomes flat to have increased disk scale length and fainter central surface brightness. 

\section{B. Summary page of the fitting result} 
\label{ap:summary}
We present a summary page of 2D fitting result of sample galaxies.  
A complete figure set for 144 galaxies are available in the online journal.


\clearpage
\begin{deluxetable}{llrrrr}
\centering
\setlength\tabcolsep{3pt}
\tablecolumns{6}
\tablenum{1}
\tabletypesize{\scriptsize}
\tablecaption{Sample
\label{tab:sample}}
\tablewidth{0pt}
\tablehead{
 \colhead{Galaxy} &
 \colhead{MIR Classification} &
 \colhead{T type} &
 \colhead{$\rm{R_{25.5}}$} &
 \colhead{$\rm{M_{3.6}}$} &
 \colhead{Distance} \\
 & & & \colhead{[arcsec]} & \colhead{ [AB mag]} & \colhead{[Mpc]} \\
 \colhead{(1)} & \colhead{(2)} & \colhead{(3)} & \colhead{(4)} & \colhead{(5)} & \colhead{(6)} }
\startdata
     ESO013-016 & $\rm{                          SB(rs)cd      }$ &  7.5 &    73.1 &   -18.7 &   23.0  \\
     ESO026-001 & $\rm{                     (R_2')SAB(s)c      }$ &  5.9 &    57.2 &   -17.9 &   19.2  \\
     ESO027-001 & $\rm{                            SB(s)b      }$ &  5.0 &   103.2 &   -19.4 &   18.3  \\
     ESO079-007 & $\rm{                           SB(s)dm      }$ &  4.0 &    55.1 &   -18.4 &   25.2  \\
     ESO404-003 & $\rm{                            SB(s)c      }$ &  3.9 &    54.2 &   -18.9 &   29.1  \\
\enddata
\tablecomments{The full catalog contains 144 objects. Only first five entries are shown here for guidance regarding its form and content. Table 1 is published in its entirety in the electronic edition.
\\
Sample galaxies drawn from \s4g.
 (1) Galaxy name. (2) MIR Morphological classification by Buta et al. (2010) and R. Buta et al. (2013, in prep). (3) Numerical morphological type from Hyperleda.
(4) Radius at $\mu_{3.6{ \mu {\rm m}}}$=25.5 AB mag from {\s4g} Pipeline 3.
(5) Total magnitude at 3.6 \mus from \s4g Pipeline 3. (6) Distance in Mpc from NED.  We made use of the mean redshift-independent distance from NED when available, otherwise, the redshift-based distance.}
\end{deluxetable}

\clearpage
{
\begin{turnpage}
\begin{deluxetable}{lrrrrrrrrrrrrrrrrrrrrrrrr}
\centering
\label{tab:res}
\tabletypesize{\scriptsize}
\setlength\tabcolsep{3pt}
\tablecolumns{25}
\tablenum{2}
\tablecaption{Structural Parameters from {\sc budda} Image Decompositions}
\tablewidth{0pc}
\tablehead{  & \multicolumn{5}{c}{Bulge} & \multicolumn{7}{c}{Disk}  & \multicolumn{6}{c}{Bar} & \multicolumn{6}{c}{Ratio}\\
 \colhead{Object} &
 \colhead{$n_{\rm bul}$} &
 \colhead{$r_{\rm eff}$}&
 \colhead{$\mu_{\rm eff}$}&
 \colhead{PA} &
 \colhead{$\epsilon$} &
 \colhead{$\mu_{\rm 0,in}$} &
 \colhead{$\mu_{\rm 0,out}$} &
 \colhead{$h_{\rm in}$} &
 \colhead{$h_{\rm out}$} &
 \colhead{$r_{\rm br}$} &
 \colhead{PA} &
 \colhead{$\epsilon$} &
 \colhead{$n_{\rm bar}$} &
 \colhead{$r_{\rm bar}$} &
 \colhead{$R_{\rm bar}$} &
 \colhead{$c$} &
 \colhead{PA} &
 \colhead{$\epsilon$} &
 \colhead{B/T} &
 \colhead{D$_{\rm i}/$T} &
 \colhead{D$_{\rm o}/$T} &
 \colhead{D/T} &
 \colhead{Bar/T} &
 \colhead{PS/T} \\
  & & \colhead{['']} & \colhead{$[m/''^2]$} & \colhead{[deg]} & &
\colhead{$[m/''^2]$} & \colhead{$[m/''^2]$} &  \colhead{['']} & \colhead{['']} & \colhead{['']} &
\colhead{[deg]} & &
 &  \colhead{['']} & \colhead{['']}& & \colhead{[deg]} & &
 & & & & & \\
\colhead{(1)} & \colhead{(2)} & \colhead{(3)} & \colhead{(4)} & \colhead{(5)} & \colhead{(6)} &\colhead{(7)} & \colhead{(8)} & \colhead{(9)} & \colhead{(10)} & \colhead{(11)} & \colhead{(12)} &
\colhead{(13)} & \colhead{(14)} & \colhead{(15)} & \colhead{(16)} & \colhead{(17)} & \colhead{(18)} &\colhead{(19)} & \colhead{(20)} & \colhead{(21)} & \colhead{(22)} & \colhead{(23)} & \colhead{(24)} & \colhead{(25)}
}
\startdata
ESO013-016  &  \nodata &  \nodata &  \nodata &  \nodata &  \nodata &  21.4 &  19.9 &  31.0 &  15.4 &   39 &   79 & 0.37 &  0.80 &   15
 &   15 &  2.7 &   77 & 0.65 &  \nodata &  0.55 &  0.38 &  0.93 &  0.07 &  \nodata \\
ESO026-001  &  \nodata &  \nodata &  \nodata &  \nodata &  \nodata &  21.6 &  19.9 &  27.9 &  12.1 &   36 &  126 & 0.13 &  1.17 &   11 &   11 &  2.7 &  158 & 0.55 &  \nodata &  0.60 &  0.30 &  0.90 &  0.10 &  \nodata \\
ESO027-001  \tablenotemark{a}   &   1.9 &   2.5 &  18.3 &  146 & 0.42 &  20.6 &  19.2 &  40.5 &  16.2 &   37 &   33 & 0.19 &  0.34 &   27 &   34 &  2.8 &  151 & 0.56 &  0.11 &  0.41 &  0.37 &  0.78 &  0.11 &  \nodata \\
ESO079-007  &  \nodata &  \nodata &  \nodata &  \nodata &  \nodata &  20.9 &  \nodata &  12.9 &  \nodata &  \nodata &   90 & 0.16 &  0.35 &   15 &   15 &  2.8 &  109 & 0.76 &  \nodata &  0.97 &  \nodata &  0.97 &  0.03 &  \nodata \\
ESO404-003  &  \nodata &  \nodata &  \nodata &  \nodata &  \nodata &  20.1 &  \nodata &  12.6 &  \nodata &  \nodata &  111 & 0.52 &  0.68 &   13 &   13 &  2.7 &  103 & 0.82 &  \nodata &  0.97 &  \nodata &  0.97 &  0.03 &  \nodata \\
\enddata
\tablenotetext{a}{Galaxies that have a lens or inner ring. Derived break radii of theses galaxies are the lengths of lens or inner ring. Inner (outer) disk scale lengths of these galaxies are disk scale length of disk inside (outside) of lens or inner ring. These galaxies do not show another break further out. See Section 4 for details.}
\tablenotetext{b}{Inner ring or lens was modeled, but the galaxy has another break further out.}
\tablenotetext{c}{Galaxies that have two breaks (TII.o + TIII). The inner break was modeled, but these are not inner rings nor lenses.}
\tablecomments{
The full catalog contains 144 objects. Only first five entries are shown here for guidance regarding its form and content.
Table 2 is published in its entirety in the electronic edition.\\
Results from the image decomposition. 
Column (1) gives the galaxy name; columns (2) to (6) correspond to the bulge component only and display, respectively, the S\'ersic index, effective radius and surface brightness, position angle and ellipticity. Columns (7) to (13) correspond to the disk components, respectively: central surface brightness of inner and outer disks, scale length of inner and outer disks, break radius, position angle and ellipticity. The bar parameters are followed. (14): S\'ersic index of bar, (15) projected length of the semi-major axis. Note that in Figure~\ref{fig:rbar_hi},\ref{fig:rbr} we plot deprojected bar radii. (16) boxiness (i.e. shape parameter $c$), (17) position angle and (18) ellipticity of the bar. Columns (19) to (24) show the fraction of the total galaxy luminosity in each model component, respectively: bulge, inner disk, outer disk, disk (i.e the sum of inner and outer disks), bar and point source. All size measures are in arcseconds and not deprojected. Intensities are in units of 3.6 \mus AB magnitudes per square arcsecond.}
\end{deluxetable}
\end{turnpage}
}

\setcounter{figure}{13}
\clearpage
\addtocounter{figure}{-1}
\begin{figure*}
\label{fig:ap_b}
\centering
\includegraphics[keepaspectratio=true,width=\textwidth,clip=true]{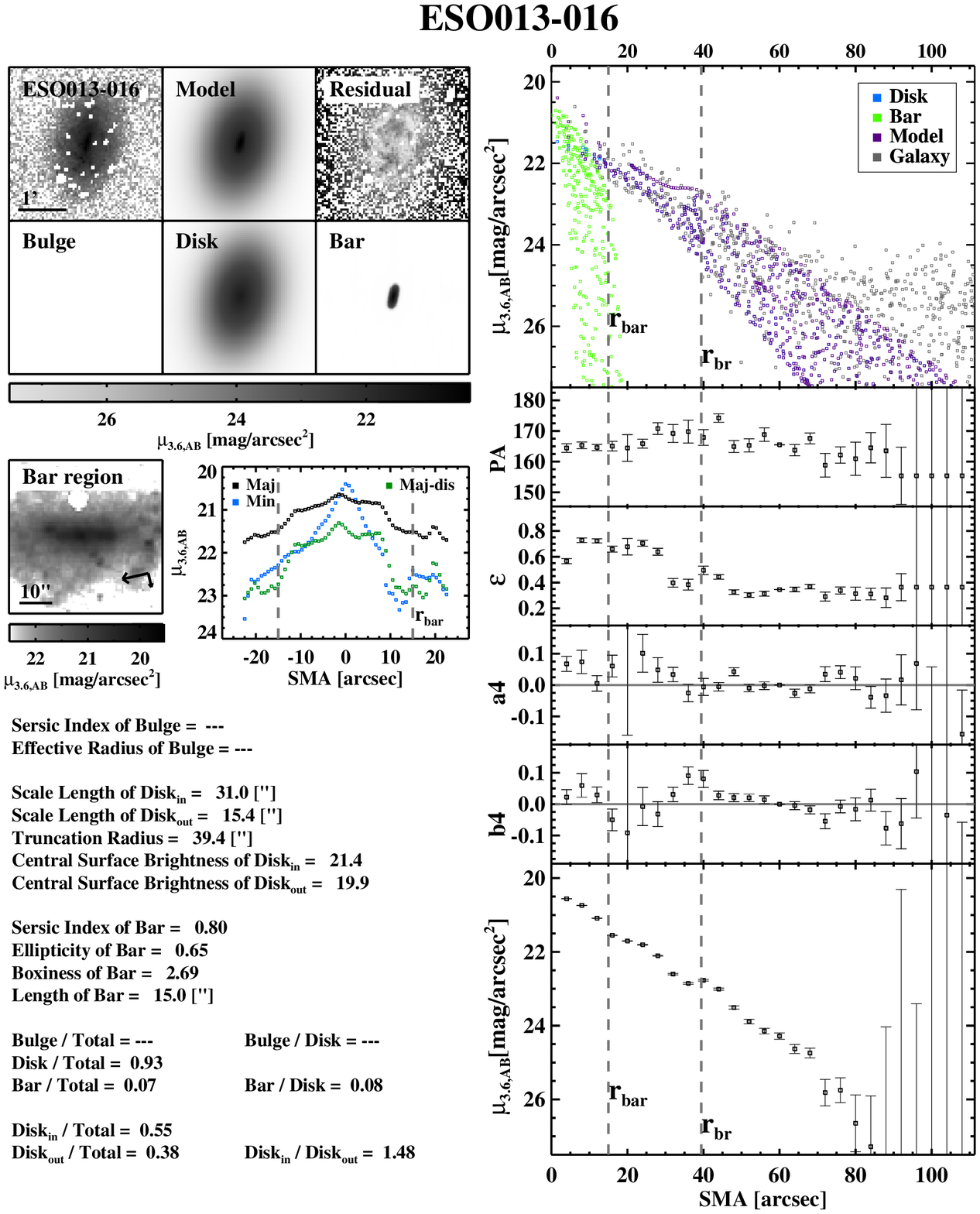}
\vskip -1.5cm
\caption{The same as Figure 4, but for ESO013-016. The full set of figures are available in the online version of the Journal.}
\end{figure*}

\end{document}